\documentclass[aps,prl,twocolumn,showpacs,amsmath,amssymb,groupedaddress]{revtex4-1}
\usepackage{graphicx}
\usepackage{color}

\usepackage[normalem]{ulem}

\def\be{\begin{equation}}
\def\ee{\end{equation}}
\def\ba{\begin{eqnarray}}
\def\ea{\end{eqnarray}}
\def\bc{\begin{center}}
\def\ec{\end{center}}

\def\rperp{r_{\!{_\perp}}}

\renewcommand{\vec}[1]{\mathbf{#1}}

\begin{document}
\title{Paraxial Theory of Direct Electro-Optic Sampling of the
Quantum Vacuum}

\author{A.S. Moskalenko}

\email{andrey.moskalenko@physik.uni-konstanz.de}

\author{C. Riek}

\author{D.V. Seletskiy}

\author{G. Burkard}

\author{A. Leitenstorfer}

\affiliation{Department of Physics and Center for Applied
Photonics, University of Konstanz, D-78457 Konstanz, Germany}

%

\date{\today}

\begin{abstract}
Direct detection of vacuum fluctuations and analysis of sub-cycle quantum properties of the
electric field are explored by a paraxial quantum theory of ultrafast electro-optic sampling.
The feasibility of such experiments is demonstrated by realistic
calculations adopting a thin ZnTe electro-optic crystal and stable
few-femtosecond laser pulses.
We show that nonlinear mixing of a short near-infrared
probe pulse with multi-terahertz vacuum field modes leads to an increase of the
signal variance with respect to the shot noise level.
The vacuum contribution increases significantly
for appropriate length of the nonlinear crystal, short probe pulse
durations, tight focusing, and sufficiently large number of
photons per probe pulse.
If the vacuum input is squeezed, the
signal variance depends on the probe delay.
Temporal positions with noise level below the pure vacuum may be traced with a sub-cycle accuracy.
\end{abstract}


\maketitle


Finite fluctuation amplitudes in the ground state of empty space
represent the ultimate hallmark of the quantum nature of the
electromagnetic radiation field.  These vacuum fluctuations
manifest themselves indirectly in a number of phenomena that are
accessible to spectroscopy such as the spontaneous decay of
excited atomic states as well as the Lamb shift
\cite{Sakurai1967} in atoms \cite{Lamb1947} and in
quantum-mechanical electric circuits \cite{Fragner2008}. Access to
the quantum aspects of electromagnetic radiation is provided by
the analysis of photon correlation
\cite{Hanbury_Brown1956,Kimble1977} or homodyning
\cite{Shapiro1979,Mandel1982,Slusher1985,Smithey1993,Breitenbach1997,Silberhorn2007}
measurements. However, these approaches require amplification of
the quantum field under study  to finite intensity and averaging
of the information over multiple optical cycles.

On the other side, precise determination of voltage or electric field amplitude as a
function of time represents a fundamental task in science and
engineering.
Optical techniques have to be applied when detecting electric
fields oscillating in the terahertz (THz) range and above. Those
approaches involve probing with ultrashort laser pulses of a
temporal duration on the order of half an oscillation period at
the highest frequencies under study. Far-infrared electric
transients \cite{Auston1984_2,Fattinger1989} can be characterized
by photoconductive switching \cite{Auston1975}.
Electro-optic sampling in free space \cite{Wu1995,Nahata1996,Gallot1999}
allows field-resolved detection at high sensitivity in the entire
far- and mid-infrared spectral range \cite{Liu2004,Kuebler2004}.
Direct studies of the complex-valued susceptibilities of materials
and the elementary dynamics in condensed matter may be performed
with these methods
%
%
\cite{Basov2011,Ulbricht2011}. The time integral of near-infrared
to visible electric-field wave packets is accessible with
attosecond streaking \cite{Kienberger2004}. So far, all those
techniques were restricted to the classical field amplitude.

In this Letter,
we demonstrate theoretically that the quantum properties of light
may be accessed directly in the time domain, i.e. with sub-cycle
temporal resolution. Our considerations are based on the realistic
example of electro-optic detection with zincblende-type materials
\cite{Planken2001}. Even vacuum fluctuations may be
sampled without amplification by broadband probing of
electric field amplitudes in the multi-THz region with
few-femtosecond laser pulses of moderate energy content.

\begin{figure}
  \includegraphics[width=1.0\linewidth]{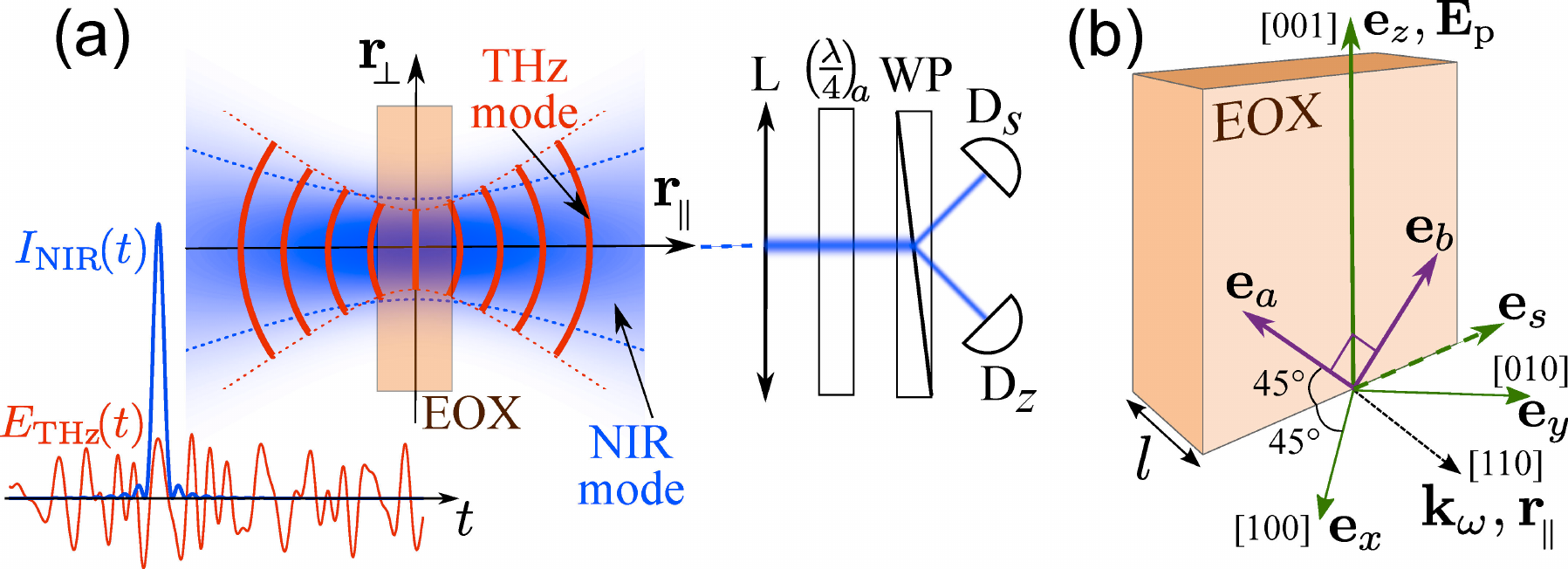}\\
  \caption{(color online) Electro-optic sampling setup and geometry.
  (a) The incoming near-infrared (NIR) probe and multi-THz signal fields mix in the electro-optic crystal (EOX).
  The NIR (blue) spatial mode amplitude is depicted by the contour plot
  whereas a sampled THz (red) spatial mode is indicated by wave fronts.
  Bottom left corner: time profiles of the NIR intensity envelope $I_{_\mathrm{NIR\!}}(t)$ and a representative multi-THz
  vacuum field $E_{_\mathrm{THz\!}}(t)$.
  After collimation by a lens (L), the modified NIR field is analyzed using
  a quarter-wave plate $\left(\!\frac{\lambda}{4}\!\right)_{\!a}$, a Wollaston prism (WP) and
  balanced detectors (D$_s$,D$_z$) measuring the difference in the photon
  flux for the split components. (b) Spatial directions determining the
  electro-optic effect in the zincblende-type EOX and the following ellipsometry analysis.
  }\label{Fig:Geometry}
\end{figure}

We consider the geometry of electro-optic sampling shown in Fig.~\ref{Fig:Geometry}.
An ultrashort near-infrared (NIR) wave packet with electric field  $\vec{E}_{\mathrm{p}}$ propagates along
the $[110]$ axis of an electro-optic crystal (EOX) \cite{Namba1961,Planken2001}.
%
%
Its wave vector
$\vec{k}_\omega$ is perpendicular to the $z$-axis $\vec{e}_z$ of
the EOX. We select $\vec{E}_{\mathrm{p}}
\parallel \vec{e}_z$ \cite{Suppl_Mat}. In this configuration, the
second-order nonlinear mixing of $\vec{E}_{\mathrm{p}}(t)$ with
an incident THz field $\hat{\vec{E}}_{_\mathrm{THz}}(t)$ induces
nonlinear polarization in the EOX plane with the components (for
details, see Ref.~\cite{Suppl_Mat})
\begin{equation}\label{Eq:Psz_t}
   \hat{P}^{(2)}_s(t)=-\epsilon_0d\hat{E}_{_\mathrm{THz},s}(t)E_{\mathrm{p}}(t), \ \ \ \hat{P}^{(2)}_z(t)=0\;.
\end{equation}
$\epsilon_0$ is the vacuum permittivity. The coupling constant
$d=-n^4 r_{_{\!41}}$ can be determined from the electro-optic
coefficient $r_{_{\!41}}$ and refractive index (RI) $n$ at the
central frequency $\omega_\mathrm{c}$ of $\vec{E}_{\mathrm{p}}$
\cite{New_book,Powers_book,Yariv_book}. In general, both fields
$\hat{E}_{_\mathrm{THz}}\equiv \hat{E}_{_\mathrm{THz},s}$
and $\hat{P}^{(2)}_s$ in
Eq.~\eqref{Eq:Psz_t} are quantized, whereas
$E_{\mathrm{p}}=E_{\mathrm{p},z}=\langle
\hat{E}_{\mathrm{p},z}\rangle$ denotes the classical part of the
probe field. We neglect the effect of quantum mechanical
fluctuations of the probe field on $\hat{\vec{P}}^{(2)}$, assuming
a sufficiently large $E_{\mathrm{p}}$.

The nonlinear polarization $\hat{\vec{P}}^{(2)}$ generated by the
wave mixing in the EOX represents a source in the inhomogeneous
wave equation describing propagation of the electric field
$\hat{\vec{E}}$ in the EOX. The fields
$\hat{\vec{F}}=\hat{\vec{E}},\hat{\vec{P}}^{(2)}$
propagating in the forward direction $r_{_\|}$ (see Fig.~\ref{Fig:Geometry})
can be decomposed as
$\hat{\vec{F}}(\vec{r},t)\!=\!\!\int_{-\infty}^\infty
\!\!\mathrm{d}\omega\  \hat{\vec{F}}(\vec{r};\omega)
e^{i(k_{\omega}r_{_\|}-\omega
    t)},$
where
$k_\omega=\omega n_\omega/c_0$. $c_0$ and $n_\omega$ are the
velocity of light and the frequency-dependent RI of the EOX,
respectively. Using the paraxial approximation
\cite{Boyd_book,Shen_book}, the inhomogeneous wave equation reads
\begin{equation}\label{Eq:wave_propagation_paraxial}
    \left[\Delta_{_\perp}+2ik_\omega\frac{\partial}{\partial
    r_{_\|}}\right]\hat{\vec{E}}(\vec{r};\omega)=-\frac{\omega^2}{\epsilon_0c_0^2}\hat{\vec{P}}^{(2)}(\vec{r};\omega)\;,
\end{equation}
where $\vec{r}_{\!{_\perp}}=(r_s,r_z)$ and
$\Delta_{_\perp}=\frac{\partial^2}{\partial
r_s^2}+\frac{\partial^2}{\partial r_z^2}$. From
Eq.~\eqref{Eq:Psz_t} we obtain
   $\hat{P}^{(2)}_s(\vec{r};\omega)=-\epsilon_0d\int_{-\infty}^\infty\!\mathrm{d}\Omega\;
   \hat{E}_{_\mathrm{THz}}(\vec{r};\Omega)
   E_{\mathrm{p}}(\vec{r};\omega-\Omega)
   e^{i(k_{\Omega}+k_{\omega-\Omega}-k_{\omega})r_{_\|}}\;.$
The electric field of the probe beam provides a solution of the
homogeneous part of Eq.~\eqref{Eq:wave_propagation_paraxial} which
can be decomposed into Laguerre-Gaussian (LG) modes
\cite{Allen1992,Calvo2006}  (see Ref.~\cite{Suppl_Mat}). We
adopt a probe pulse train with a fundamental Gaussian transverse mode of amplitude
$\alpha_{\mathrm{p}}(\omega)$:
\begin{equation}\label{Eq:Ep_r_omega}
   E_{\mathrm{p}}(\vec{r};\omega)=\alpha_{\mathrm{p}}(\omega)\mathrm{LG}_{_{00}}(\vec{r}_{\!{_\perp}},r_{_\|};k_\omega)\;.
\end{equation}
A length $l$ of the EOX much shorter than the Rayleigh range of a
beam at the relevant THz frequencies $\Omega$ with waist size
$w_0$ is assumed, i.e. $l\ll k_\Omega w_0^2/2$.

The EOX is located at the beam waist, $r_{_\|}=0$, and has
anti-reflection coating on its surfaces. Denoting
$\hat{\vec{F}}(\vec{r}_{\!{_\perp}};\omega)\equiv
\hat{\vec{F}}(\vec{r}_{\!{_\perp}},r_{_\|}\!=\!0;\omega)$ we find
that at the exit from the EOX, $r_{_\|}=l/2$, the total electric
field in the $(110)$ plane is
given by
\begin{equation}\label{Eq:E_EOX_exit}
  \hat{\vec{E}}'(\boldsymbol{\mathcal{Y}})=E_\mathrm{p}(\boldsymbol{\mathcal{Y}})\vec{e}_z+\hat{E}^{(2)}(\boldsymbol{\mathcal{Y}})\vec{e}_s
  +\delta\hat{\vec{E}}'(\boldsymbol{\mathcal{Y}})\;,
\end{equation}
where $\boldsymbol{\mathcal{Y}}\equiv \{\vec{r}_{\!{_\perp}};\omega\}$. $\delta
\hat{\vec{E}}'(\boldsymbol{\mathcal{Y}})=\hat{\vec{E}}_p(\boldsymbol{\mathcal{Y}})-E_p(\boldsymbol{\mathcal{Y}})
\vec{e}_z$ denotes the vacuum field contribution at the probe
frequency $\omega$ in the vacuum picture \cite{Knight_book}.
The correction to the probe field generated in the EOX is
evaluated as
\begin{equation}\label{Eq:E2_exit}
  \begin{split}
  \hat{E}^{(2)}(\vec{r}_{\!{_\perp}}\!;\omega)\!=\!\!
  \int_{-\infty}^\infty\!\!\!\!\!\mathrm{d}\Omega\ \hat{E}_{_\mathrm{THz}}(\vec{r}_{\!{_\perp}}\!;\Omega)
   E_{\mathrm{p}}(\vec{r}_{\!{_\perp}};\omega\!-\!\Omega) \zeta_{\omega,\Omega}\;,
   \end{split}
\end{equation}
where the factor
$\zeta_{\omega,\Omega}=-id\frac{l \omega}{2c_0 n }
   \mathrm{sinc}\!\left[\frac{l\Omega}{2c_0}(n_{_\Omega}-n_\mathrm{g})\right]$
determines phase matching. Here $\mathrm{sinc}(x)\equiv
\sin(x)/x$, $n_{_\Omega}$ is the RI at $\Omega$, whereas $n$ and
$n_\mathrm{g}$ are the RI and the group RI $c_0\partial
k_\omega/\partial \omega$ at $\omega=\omega_\mathrm{c}$,
respectively. Going beyond Ref.~\onlinecite{Gallot1999} where an
expression similar to Eq.~\eqref{Eq:E2_exit} was derived for the
case of plane waves in order to establish a classical theory of
electro-optic sampling, Eqs.~\eqref{Eq:E_EOX_exit} and
\eqref{Eq:E2_exit} include the transverse spatial dependence of
the fields, the quantized form of the signal as well as the contribution of quantum fluctuations at the probe frequencies.
These points are crucial for our further analysis.

From Eq.~\eqref{Eq:E_EOX_exit}, we see that the nonlinear mixing of the
probe and THz components generates a new field
propagating in the same direction and polarized perpendicular to
the probe. For the analysis of the polarization state of the modified probe, we consider the
field components in the coordinate frame
$\vec{e}_{\substack{a\\b}} = (\vec{e}_z\mp \vec{e}_s )/\sqrt{2}$
rotated by $45^\circ$ with respect to the $\vec{e}_{z,s}$ frame
[Fig.~\ref{Fig:Geometry}(b)],
   $\hat{E}'_{\substack{a\\[0pt]b}}(\boldsymbol{\mathcal{Y}})=E_{\mathrm{p}}(\boldsymbol{\mathcal{Y}})\big[1\pm
   i\hat{\phi}(\boldsymbol{\mathcal{Y}})\big]/\sqrt{2}
   +\delta \hat{E}'_{\substack{a\\[0pt]b}}(\boldsymbol{\mathcal{Y}})\;.$
Here $\hat{\phi}(\boldsymbol{\mathcal{Y}})=i
   \hat{E}^{(2)}(\boldsymbol{\mathcal{Y}})/E_{\mathrm{p}}(\boldsymbol{\mathcal{Y}})\;$
must be small for the frequency range of the probe.

The ellipsometry setup used in typical experiments is explained in
Fig.~\ref{Fig:Geometry}(a). We consider its effects at the exit
surface of the EOX. This simplification is justified when all
probe photons are detected without spatial filtering.
The first step of the analysis consists in describing the action
of the quarter-wave plate with axes oriented along $\vec{e}_a$ and
$\vec{e}_b$ such that it phase-shifts the $a$-component of the
field by $\pi/2$: $\hat{E}''_{a}(\boldsymbol{\mathcal{Y}})=i
\hat{E}'_{a}(\boldsymbol{\mathcal{Y}})$,
$\hat{E}''_{b}(\boldsymbol{\mathcal{Y}})=\hat{E}'_{b}(\boldsymbol{\mathcal{Y}})$. The
Wollaston prism splits the electric field into its $z$- and
$s$-components:
\begin{equation}\label{Eq:E_prime2_zs}
   \hat{E}''_{\substack{z\\[-1pt]s}}(\boldsymbol{\mathcal{Y}})=
   \frac{e^{\pm i\frac{\pi}{4}}E_{\mathrm{p}}(\boldsymbol{\mathcal{Y}})}{\sqrt{2}}\big[1\mp \hat{\phi}(\boldsymbol{\mathcal{Y}})\big]
   +\delta \hat{E}''_{\substack{z\\[-1pt]s}}(\boldsymbol{\mathcal{Y}})\;.
\end{equation}
Finally, the photon numbers in both field components are detected
and subtracted. The photon number operator for the polarization
$\alpha=z,s$ reads \footnote{Confer Ref.~\cite{Raymer1995} for a
case with a simpler transverse mode structure}
\begin{equation}\label{Eq:Nz}
   \hat{\mathcal{N}}_\alpha=C\int_0^\infty\!\!\!\mathrm{d}\omega \frac{\eta(\omega)}{\hslash\omega}
   \int\!\mathrm{d}^2\rperp
   \hat{E}''^\dagger_\alpha(\vec{r}_{\!{_\perp}}\!;\omega)
   \hat{E}''_\alpha(\vec{r}_{\!{_\perp}}\!;\omega)\;,
\end{equation}
where $C=4\pi c_0 n \epsilon_0$, the dagger denotes Hermitian
conjugation and the spatial integral covers the entire transverse
profile of the probe beam. The frequency-dependent quantum
efficiency of the photodetector $\eta(\omega)\approx 1$ over the
detected frequency range but vanishes quickly for
$\omega\rightarrow 0$.

Inserting Eq.~\eqref{Eq:E_prime2_zs} into Eq.~\eqref{Eq:Nz} and
neglecting the second-order terms in $\delta \hat{\vec{E}}''$ as well
as the mixed terms depending linearly both on $\delta
\hat{\vec{E}}''$ and on $\hat{E}_{_\mathrm{THz}}$ (contained in
$\hat{\phi}$) \footnote{The mixed terms were also neglected
already in Eq.~\eqref{Eq:Psz_t}. The second order terms in $\delta
\hat{\vec{E}}''$ ($\delta \hat{E}''^\dagger_z\delta \hat{E}''_z$
and $\delta \hat{E}''^\dagger_s\delta \hat{E}''_s$) do not
contribute to the expectation value of the signal, neither to its
variance or any higher moments.},
%
%
we obtain for the total detected quantum signal
\begin{equation}\label{Eq:S}
   \hat{\mathcal{S}}\equiv\hat{\mathcal{N}}_s-\hat{\mathcal{N}}_z=\hat{\mathcal{S}}_\mathrm{eo}+\hat{\mathcal{S}}_\mathrm{sn}\;,
\end{equation}
where the electro-optic signal (EOS)
$\hat{\mathcal{S}}_\mathrm{eo}$ is
\begin{equation}
   \hat{\mathcal{S}}_\mathrm{eo} = C\! \int\!\!\mathrm{d}^2 \rperp\!\!
\int_0^\infty\!\!\!\!\!\mathrm{d}\omega
\frac{\eta(\omega)}{\hslash\omega}\;
|E_{\mathrm{p}}(\boldsymbol{\mathcal{Y}})|^2
\!\left[\hat{\phi}(\boldsymbol{\mathcal{Y}})+\mathrm{H.c.}\right]\!
\label{Eq:S_eo}
\end{equation}
and the shot noise (SN) contribution
$\hat{\mathcal{S}}_\mathrm{sn}$ reads
$ \hat{\mathcal{S}}_\mathrm{sn} =  C\! \int\!\mathrm{d}^2
\rperp\! \int_0^\infty\!\mathrm{d}\omega
\frac{\eta(\omega)}{\hslash\omega}
\!\left[E_{\mathrm{p}}^*(\boldsymbol{\mathcal{Y}})\delta
\hat{E}''_+(\boldsymbol{\mathcal{Y}})+\mathrm{H.c.}\right]\!.$
%
Here $\mathrm{H.c.}$ denotes the Hermitian conjugate and $\delta
\hat{E}''_+(\boldsymbol{\mathcal{Y}})=e^{i\pi/4}[\delta
\hat{E}''_s(\boldsymbol{\mathcal{Y}})+i\delta \hat{E}''_z(\boldsymbol{\mathcal{Y}})]/\sqrt{2}$
is the circular component of the probe field vacuum contribution
\footnote{The phase shift is of no physical importance for the
vacuum field contribution}.  Summing up the signals from
both detectors, we obtain the expectation value of the number of
detected photons per probe pulse $N=\langle
\hat{\mathcal{N}}_s+\hat{\mathcal{N}}_z \rangle=\frac{4\pi c_0 n
\epsilon_0}{\hslash}\int_0^\infty\!\mathrm{d}\omega
\frac{\eta(\omega)}{\omega}|\alpha_\mathrm{p}(\omega)|^2$.

Using Eqs.~\eqref{Eq:Ep_r_omega} and \eqref{Eq:E2_exit} in
Eq.~\eqref{Eq:S_eo}, we obtain
\begin{equation}\label{Eq:S_EO2}
  \hat{\mathcal{S}}_\mathrm{eo}\!=\!\frac{dlN\omega_\mathrm{p}}{c_0n}\!\!\int\!\!\mathrm{d}^2 \rperp\;
\!g_{_{00}}^2(\vec{r}_{\!{_\perp}})\!
\int_{-\infty}^\infty\!\!\!\!\!\!\mathrm{d}\Omega\,
\hat{E}_{_\mathrm{THz}}\!(\vec{r}_{\!{_\perp}}\!;\Omega)R(\Omega).
\end{equation}
$g_{_{00}}(\vec{r}_{\!{_\perp}})\!\equiv\!\mathrm{LG}_{_{00}}(\vec{r}_{\!{_\perp}},r_{_\|}=0;k_\omega)=\sqrt{2/\pi}\,w_0^{-1}
\exp(-r_{\!{_\perp}}^2/w_0^2)$ is a normalized Gaussian independent of $\omega$
and
$\omega_\mathrm{p}=\int_0^\infty\!\mathrm{d}\omega\;
\eta(\omega)|\alpha_\mathrm{p}(\omega)|^2\big/\int_0^\infty\!\mathrm{d}\omega
\frac{\eta(\omega)}{\omega}|\alpha_\mathrm{p}(\omega)|^2$ is the
average detected frequency.
We have introduced the response function
  $R(\Omega)=\mathrm{sinc}\!\left[\frac{l\Omega}{2c_0}(n_{_\Omega}-n_\mathrm{g})\right]f(\Omega)$
with the normalized
Hermitian spectral autocorrelation function
$f(\Omega)=\big[f_{_+}^*(\Omega)+f_{_-}(\Omega)\big]/2$, where
$f_{_\pm}(\Omega)={\int_0^\infty\!\mathrm{d}\omega\;
\eta(\omega)\;\alpha_\mathrm{p}^*(\omega)\alpha_\mathrm{p}(\omega\pm\Omega)\Big/}{\int_0^\infty\!\mathrm{d}\omega\;
\eta(\omega)|\alpha_\mathrm{p}(\omega)|^2}$.

Within the paraxial
quantization \cite{Calvo2006},
$\hat{E}_{_\mathrm{THz}}(\vec{r}_{\!{_\perp}};\Omega)$ in
Eq.~\eqref{Eq:S_EO2}  is given by \cite{Suppl_Mat}
\begin{equation}\label{Eq:THz_amplitude}
  \hat{E}_{_\mathrm{THz}}(\vec{r}_{\!{_\perp}};\Omega)\!=
  -i\sum_{l,p}\sqrt{\frac{\hslash\Omega}{4\pi\epsilon_0
  c_0n_{_\Omega}}}\hat{a}_{s,l,p}(\Omega)g'_{_{lp}}(\vec{r}_{\!{_\perp}})
\end{equation}
for $\Omega>0$, $\hat{E}_{_\mathrm{THz}}(\vec{r}_{\!{_\perp}};\Omega<0)
=\hat{E}^\dagger_{_\mathrm{THz}}(\vec{r}_{\!{_\perp}};-\Omega)$. Here, $\hat{a}_{s,l,p}(\Omega)$ annihilates a
photon with frequency $\Omega$, orbital quantum numbers $l,p$ and
polarization $\vec{e}_s$. We have introduced the transverse mode
functions
$g'_{_{lp}}(\vec{r}_{\!{_\perp}})\equiv\mathrm{LG}_{_{lp}}(\vec{r}_{\!{_\perp}},r_{_\|}=0;k_\Omega)$.
In contrast to the probe beam, the waist size $w_0'$
characterizing these mode functions is a free parameter of the
expansion \eqref{Eq:THz_amplitude}. Inserting
Eq.~\eqref{Eq:THz_amplitude} into Eq.~\eqref{Eq:S_EO2} and
selecting $w_0'=w_0/\sqrt{2}$,
we can perform the
spatial integration using $\int\mathrm{d}^2
r_{_{\!\perp}}g_{_{00}}^2(\vec{r}_{_\perp})
g'_{_{lp}}(\vec{r}_{_\perp})=\frac{1}{\sqrt{\pi}w_0}\delta_{_{l,0}}\delta_{_{p,0}}$.
Then we obtain from Eq.~\eqref{Eq:S_EO2}
\begin{equation}\label{Eq:signal_eo_final}
  \hat{\mathcal{S}}_\mathrm{eo}=-i\sqrt{B}
   \int_{0}^\infty\!\!\mathrm{d}\Omega\, \sqrt{\frac{\Omega}{n_{_\Omega}}}\big[\hat{a}_{s,0,0}(\Omega) R(\Omega)
  -\mathrm{H.c.}\big],
\end{equation}
where
$B={(d^2l^2N^2\omega_\mathrm{p}^2\hslash)\big/}{(4\pi^2\epsilon_0c_0^3n^2w_0^2)}$.





As an input, we now consider a THz quantum field with no
coherent (classical) contribution: $\langle
\hat{E}_{_\mathrm{THz}} \rangle=0$, e.g., a bare multi-THz vacuum.
%
%
Then $\langle\hat{\mathcal{S}}\rangle =0$ since
$\langle\hat{\mathcal{S}}_\mathrm{sn}\rangle =0$ and
$\hat{\phi}$ in Eq.~\eqref{Eq:S_eo} depends linearly on
$\hat{E}_{_\mathrm{THz}}$, thus also
$\langle\hat{\mathcal{S}}_\mathrm{eo}\rangle =0$. However, the
variance of the signal does not vanish. If
the range of detected THz frequencies,
determined by $R(\Omega)$, does not overlap with the frequency content
of the probe
beam, the signal variance
$\langle\hat{\mathcal{S}}^2\rangle-\langle\hat{\mathcal{S}}\rangle^2=\langle\hat{\mathcal{S}}^2\rangle$
can be written as $\langle\hat{\mathcal{S}}^2\rangle
=\langle\hat{\mathcal{S}}^2_\mathrm{eo}\rangle+\langle\hat{\mathcal{S}}^2_\mathrm{sn}\rangle$.
Calculating the SN contribution
using the paraxial quantization \cite{Calvo2006}, we obtain the
expected result $\langle
\hat{\mathcal{S}}^2_{\mathrm{sn}}\rangle=N$.

Evaluating $  \langle\hat{\mathcal{S}}_\mathrm{eo}^2\rangle$
for the multi-THz vacuum yields
\begin{equation}\label{Eq:variance_THz_final}
  \langle\hat{\mathcal{S}}_\mathrm{eo}^2\rangle\!=\!N^2\left(\!n^3\frac{l \omega_\mathrm{p}}{c_0}r_{_{\!41}}\!\!\right)^{\!\!2}\,
  \frac{\hslash \int_{0}^\infty\!\mathrm{d}\Omega\; \Omega\,
                                         (n/n_{_\Omega}) |R(\Omega)|^2}{4\pi^2\epsilon_0c_0n w_0^2}
\end{equation}
where we have
used ${\langle
\hat{a}_{s,0,0}(\Omega)\hat{a}_{s,0,0}^\dagger(\Omega')\rangle}={\delta(\Omega-\Omega')}$,
whereas the expectation values of other possible quadratic
combinations of $\hat{a}_{s,l,p}^\dagger$ and $\hat{a}_{s,l,p}$
vanish. Note that the second and third factors on the right-hand
side of Eq.~\eqref{Eq:variance_THz_final} have the dimensions
(m/V)$^2$ and (V/m)$^2$, respectively. The latter can be
interpreted as the square of the effective multi-THz rms (root
mean square) vacuum electric field filtered by the response
function. The former, $\propto r_{_{41}}^2$, determines how
effectively this field is sampled for a fixed $N$.

\begin{figure}
  \includegraphics[width=1.0\linewidth]{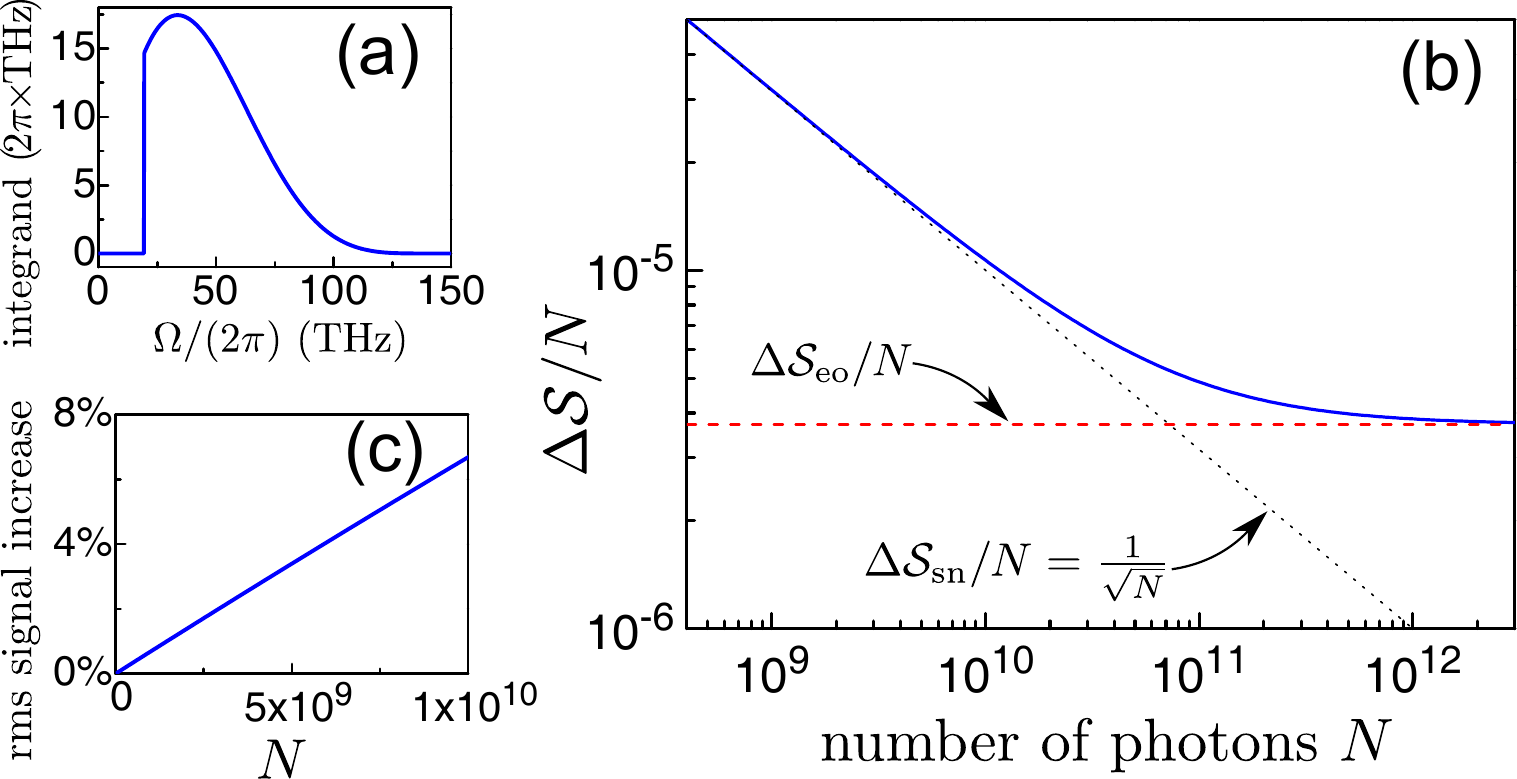}\\
  \caption{(color online) (a) Calculated integrand function $\Omega\,
    (n/n_{_\Omega}) |R(\Omega)|^2$ entering Eq.~\eqref{Eq:variance_THz_final}. (b) Double-logarithmic plot of the
  ratio $\Delta \mathcal{S}/N$ in dependence on $N$.
Black dotted  (red dashed) line shows the bare SN (multi-THz
vacuum) contribution.
  (c)
  Increase of
  $(\Delta\mathcal{S}-\Delta\mathcal{S}_{\mathrm{sn}})/\Delta\mathcal{S}_{\mathrm{sn}}$
  with $N$.
Parameters are defined in the main text.}\label{Fig:signal_vacuum}
\end{figure}

To illustrate the results,
we assume the following realistic specifications of the sampling
few-femtosecond NIR laser pulse: center frequency 255 THz,
spectral bandwidth 150 THz with rectangular spectral shape and
flat phase, leading to $\omega_{\mathrm p}=247$~THz, and waist
size $w_0=3~\mu$m \cite{Brida2014}. We consider a $l=7~\mu$m thick
ZnTe EOX with $r_{_{41}}=4$~pm/V
\cite{Cingolani1981,Leitenstorfer1999}, $n=2.76$,
$n_{\mathrm{g}}=2.9$, and $n_{_\Omega}$ varying  only slightly
(from 2.55 to 2.59) for relevant THz frequencies \cite{Suppl_Mat}.
The resulting integrand function entering
Eq.~\eqref{Eq:variance_THz_final} is shown in
Fig.~\ref{Fig:signal_vacuum}(a) (for details, see
Ref.~\cite{Suppl_Mat}). Diffraction effects are taken into account
by excluding wavelengths $\lambda$ with
$\lambda/(2n_{_\Omega})>w_0$.

Based on this input, we calculate the dependence of the rms value
of the signal $\Delta \mathcal{S}=\langle
\hat{\mathcal{S}}^2\rangle^{1/2}$ on the average number $N$ of
photons in the sampling NIR pulse, as shown in
Fig.~\ref{Fig:signal_vacuum}(b) on a double-logarithmic scale.
Above a certain $N$, the EOS contribution of the multi-THz vacuum
changes the typical SN scaling. The relative increase of the rms
value of the signal with respect to the SN level,
$(\Delta\mathcal{S}-\Delta\mathcal{S}_{\mathrm{sn}})/\Delta\mathcal{S}_{\mathrm{sn}}$,
is depicted in Fig.~\ref{Fig:signal_vacuum}(c) for moderate $N$ and with
linear scaling.  For even higher $N$, the  vacuum contribution starts
to dominate so that the dependence saturates to the constant EOS
level [Fig.~\ref{Fig:signal_vacuum}(b)]. Subtracting the SN
contribution from the total signal variance, the bare EOS variance
induced by the sampled quantum field can be obtained and analyzed.

To elaborate on this point, we apply our theory to a multi-THz vacuum which is
squeezed in an interval around a central frequency
$\Omega_\mathrm{c}$. The corresponding state of light is obtained
by applying the continuum squeezing operator
\cite{Barnett_book,Vogel_book,Blow1990}
$\exp\left[\frac{1}{2}\int_0^{2\Omega_\mathrm{c}}\mathrm{d}\Omega
\sum_{\alpha
lp}\big(\xi^*_{_\Omega}\hat{a}_{\alpha,l,p}(2\Omega_\mathrm{c}-\Omega)\hat{a}_{\alpha,l,p}(\Omega)-\mathrm{H.c.}\big)\right]$
to the multi-THz pure vacuum (PV) state considered above. Here
the frequency-dependent squeezing parameter $\xi(\Omega)$
satisfies the condition
$\xi(\Omega)=\xi(2\Omega_\mathrm{c}-\Omega)$. We assume that
all spatial and polarization modes are squeezed equally.
In this case, the EOS can be obtained from
Eq.~\eqref{Eq:signal_eo_final} applying the transformation
$\hat{a}_{s,0,0}(\Omega)\rightarrow \hat{a}_{s,0,0}(\Omega)\cosh
r_{_\Omega}-\hat{a}^\dagger_{s,0,0}(2\Omega_\mathrm{c}-\Omega)e^{i\theta_{_\Omega}}\sinh
r_{_\Omega}$ \cite{Barnett_book,Vogel_book,Blow1990} and working
in the vacuum picture. The expectation value of the signal remains zero.
Evaluation of the EOS variance for the squeezed vacuum (SV),
$\langle\hat{\mathcal{S}}_\mathrm{eo}^2\rangle_\mathrm{sv}(\tau)$,
is analogous to the PV case. However, the SV EOS variance depends
on the time delay $\tau$ of the NIR probe pulse leading to the
transformation $R(\Omega)\rightarrow R(\Omega)e^{-i\Omega\tau}$ of
the response function, a fact that was unimportant for handling the PV
[cf. Eq.~\eqref{Eq:variance_THz_final}]. For a probe pulse
symmetric with respect to $t=\tau$, i.e.
$E_\mathrm{p}(t-\tau)=E_\mathrm{p}(\tau-t)$, we find
$R(\Omega)=R_0(\Omega)e^{-i\Omega\tau}$, where $R_0(\Omega)$ is
real-valued.

\begin{figure}
  \includegraphics[width=1.0\linewidth]{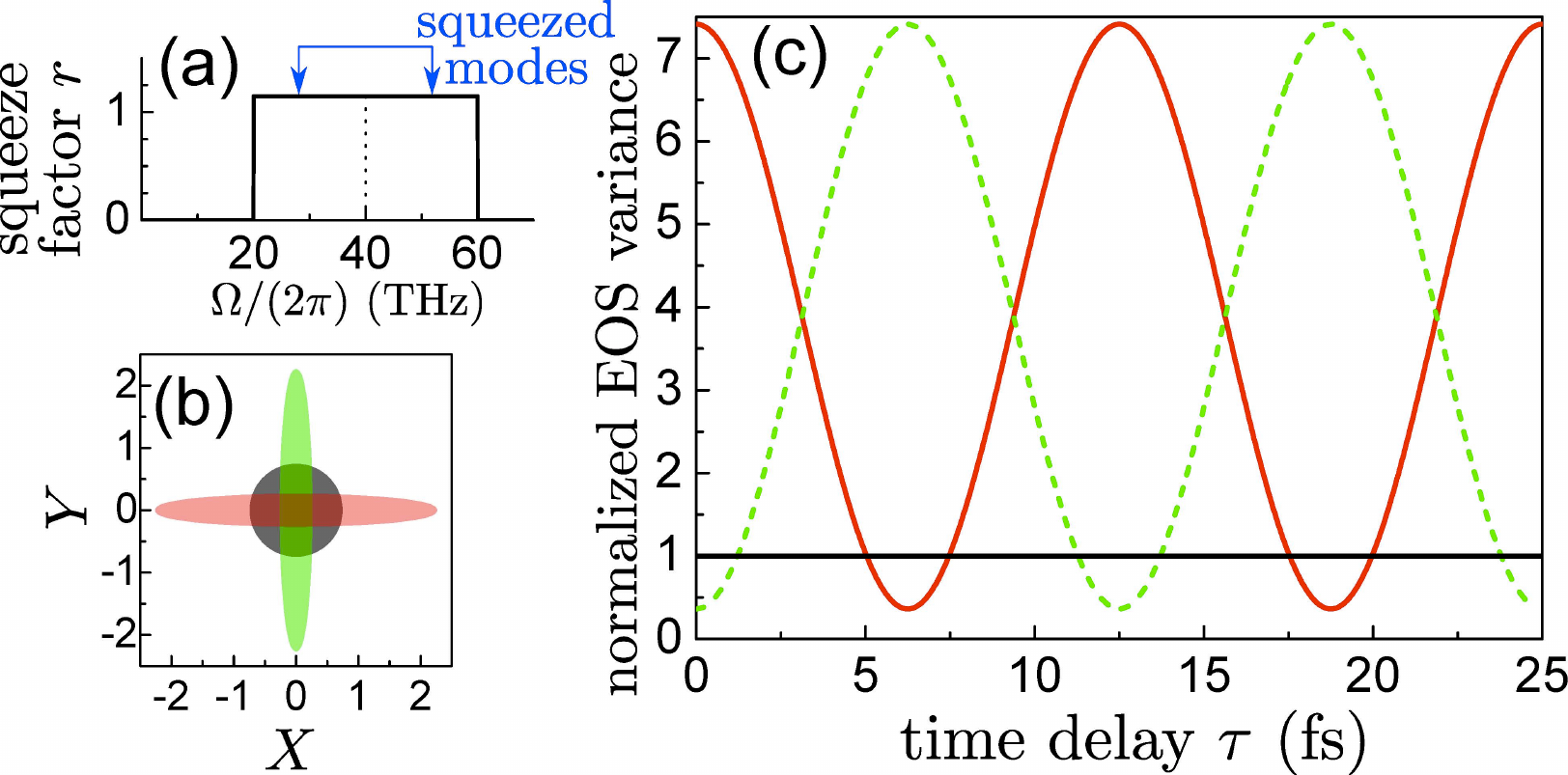}\\
  \caption{(color online) (a) Frequency dependence of the squeeze factor $r$.
  Squeezing correlates $\Omega$ and $2\Omega_\mathrm{c}-\Omega$ modes (as indicated by arrows).
  (b) Error contour in the complex-amplitude plane for PV (gray circle) and
  SV
  with $\theta\!=\!0/\theta\!=\!\pi$ (red/green ellipse with
  reduced uncertainty in the phase/amplitude quadrature $Y$/$X$).
  (c) Normalized (with respect to the constant PV level,
  solid black line) EOS variance
  in dependence on the time delay
  $\tau$ of the probe NIR pulse for SV with $\theta\!=\!0/\theta\!=\!\pi$
  (solid red line\;/\;dashed green line).
  }\label{Fig:squeezing}
\end{figure}

For our illustration we assume constant squeezing with
$\xi(\Omega)\equiv\xi=r e^{i\theta} $ in the frequency range
$[\Omega_1,\Omega_2]$ with
$\Omega_\mathrm{c}=(\Omega_1+\Omega_2)/2$, where $r=|\xi|$ is the
squeeze factor \cite{Caves1981,Walls_book} and
$\theta=\mathrm{Arg}(\xi)$ is the squeezing phase
\cite{Vogel_book}. No squeezing occurs outside this range. In
particular, we use $\Omega_\mathrm{c}/(2\pi)=40~$THz,
$\Omega_2-\Omega_1=\Omega_\mathrm{c}$ and
$\sinh r=2$ [see Fig.~\ref{Fig:squeezing}(a)].
Generalized quadrature operators \cite{Barnett_book}
$\hat{X}_\lambda=\frac{1}{\sqrt{2(\Omega_2-\Omega_1)}}\int_{\Omega_1}^{\Omega_2}\!\mathrm{d}\Omega\left[\hat{a}_{s,0,0}(\Omega)e^{-i\lambda}+\mathrm{H.c.}\right]$,
with $\hat{X}=\hat{X}_{0}$ and $\hat{Y}=\hat{X}_{\pi/2}$,
normalized so that $[\hat{X},\hat{Y}]=i$ are introduced. The error
contours for PV as well as for SV as described above and two
different squeezing phases, $\theta=0$ and $\theta=\pi$, are featured
in Fig.~\ref{Fig:squeezing}(b). The dependence of the normalized
EOS variance
$\langle\hat{\mathcal{S}}_\mathrm{eo}^2\rangle_\mathrm{sv}(\tau)/\langle\hat{\mathcal{S}}_\mathrm{eo}^2\rangle$,
where  $\langle\hat{\mathcal{S}}_\mathrm{eo}^2\rangle$ is given by
Eq.~\eqref{Eq:variance_THz_final}, on the time delay $\tau$ is
shown in Fig.~\ref{Fig:squeezing}(c) for the same states as in
Fig.~\ref{Fig:squeezing}(b) and sampling parameters used for the
PV case. For specific delay times, the EOS
variance of the multi-THz SV is by 64\% lower than the unsqueezed value of the
PV state (for details, see Ref.~\cite{Suppl_Mat}).

We emphasize the cardinal difference between our findings and
similar-looking results obtained in the context of homodyning
\cite{Anderson1997,Slusher1987}.
In homodyning experiments, the signal is determined by the
temporal overlap integral of the \textit{complex amplitudes} of
the electric fields of an input state and a local oscillator, i.e.
the information is essentially averaged over multiple oscillation
cycles of light. A restricted frequency bandwidth has to be assumed to
justify the slowly varying amplitude approximation underlying this approach.
In contrast, electro-optic sampling
provides a true sub-cycle resolution of the probed multi-THz
\textit{electric field}.
Moreover,
registration of photons is transferred
into the NIR, circumventing the lack of
efficient single-photon detectors in the multi-THz frequency
range. Most importantly,
the multi-THz quantum field may be studied without the necessity to reduce or
amplify its photon content - even if it remains in its ground state.

In conclusion, we theoretically clarify
the contribution of the quantum fluctuations of the multi-THz
vacuum electric field to the signal in ultrabroadband
electro-optic sampling by differentiating
it from the trivial shot noise of the high-frequency gating pulse.
The crucial aspects are a strong localization of the sampling beam in space and time as it
passes the nonlinear crystal, a large second-order nonlinear
coefficient and proper phase matching that might be further
optimized selecting an even more appropriate material
than the thin piece of ZnTe we have considered as an example. For a
multi-THz squeezed vacuum, the possibility to trace the oscillations of the EOS variance with
the time delay of the probe pulse is predicted. Positions occur where the noise remains significantly
below the level of unsqueezed vacuum.
The same formalism can be applied for the analysis of more complex
quantum fields in a time-resolved and non-destructive manner.
Experimental implementation of these ideas might open up a new
chapter of quantum optics operating predominantly in the time domain
and with access to sub-cycle information on the quantum state
of electromagnetic radiation.



\begin{acknowledgments}
We acknowledge funding from the ERC via the Advanced Grant 290876 ``UltraPhase'' and by DFG within SFB 767. We thank M. Kira
for useful discussions.
\end{acknowledgments}

%
%


\begin{thebibliography}{47}%
\makeatletter
\providecommand \@ifxundefined [1]{%
 \@ifx{#1\undefined}
}%
\providecommand \@ifnum [1]{%
 \ifnum #1\expandafter \@firstoftwo
 \else \expandafter \@secondoftwo
 \fi
}%
\providecommand \@ifx [1]{%
 \ifx #1\expandafter \@firstoftwo
 \else \expandafter \@secondoftwo
 \fi
}%
\providecommand \natexlab [1]{#1}%
\providecommand \enquote  [1]{``#1''}%
\providecommand \bibnamefont  [1]{#1}%
\providecommand \bibfnamefont [1]{#1}%
\providecommand \citenamefont [1]{#1}%
\providecommand \href@noop [0]{\@secondoftwo}%
\providecommand \href [0]{\begingroup \@sanitize@url \@href}%
\providecommand \@href[1]{\@@startlink{#1}\@@href}%
\providecommand \@@href[1]{\endgroup#1\@@endlink}%
\providecommand \@sanitize@url [0]{\catcode `\\12\catcode `\$12\catcode
  `\&12\catcode `\#12\catcode `\^12\catcode `\_12\catcode `\%12\relax}%
\providecommand \@@startlink[1]{}%
\providecommand \@@endlink[0]{}%
\providecommand \url  [0]{\begingroup\@sanitize@url \@url }%
\providecommand \@url [1]{\endgroup\@href {#1}{\urlprefix }}%
\providecommand \urlprefix  [0]{URL }%
\providecommand \Eprint [0]{\href }%
\providecommand \doibase [0]{http://dx.doi.org/}%
\providecommand \selectlanguage [0]{\@gobble}%
\providecommand \bibinfo  [0]{\@secondoftwo}%
\providecommand \bibfield  [0]{\@secondoftwo}%
\providecommand \translation [1]{[#1]}%
\providecommand \BibitemOpen [0]{}%
\providecommand \bibitemStop [0]{}%
\providecommand \bibitemNoStop [0]{.\EOS\space}%
\providecommand \EOS [0]{\spacefactor3000\relax}%
\providecommand \BibitemShut  [1]{\csname bibitem#1\endcsname}%
\let\auto@bib@innerbib\@empty
\bibitem [{\citenamefont {Sakurai}(1967)}]{Sakurai1967}%
  \BibitemOpen
  \bibfield  {author} {\bibinfo {author} {\bibfnamefont {J.~J.}\ \bibnamefont
  {Sakurai}},\ }\href@noop {} {\emph {\bibinfo {title} {Advanced Quantum
  Mechanics}}}\ (\bibinfo  {publisher} {Addison-Wesley},\ \bibinfo {year}
  {1967})\BibitemShut {NoStop}%
\bibitem [{\citenamefont {Lamb}\ and\ \citenamefont
  {Retherford}(1947)}]{Lamb1947}%
  \BibitemOpen
  \bibfield  {author} {\bibinfo {author} {\bibfnamefont {W.~E.}\ \bibnamefont
  {Lamb}}\ and\ \bibinfo {author} {\bibfnamefont {R.~C.}\ \bibnamefont
  {Retherford}},\ }\href {\doibase 10.1103/PhysRev.72.241} {\bibfield
  {journal} {\bibinfo  {journal} {Phys. Rev.}\ }\textbf {\bibinfo {volume}
  {72}},\ \bibinfo {pages} {241} (\bibinfo {year} {1947})}\BibitemShut
  {NoStop}%
\bibitem [{\citenamefont {Fragner}\ \emph {et~al.}(2008)\citenamefont
  {Fragner}, \citenamefont {Goppl}, \citenamefont {Fink}, \citenamefont {Baur},
  \citenamefont {Bianchetti}, \citenamefont {Leek}, \citenamefont {Blais},\
  and\ \citenamefont {Wallraff}}]{Fragner2008}%
  \BibitemOpen
  \bibfield  {author} {\bibinfo {author} {\bibfnamefont {A.}~\bibnamefont
  {Fragner}}, \bibinfo {author} {\bibfnamefont {M.}~\bibnamefont {Goppl}},
  \bibinfo {author} {\bibfnamefont {J.~M.}\ \bibnamefont {Fink}}, \bibinfo
  {author} {\bibfnamefont {M.}~\bibnamefont {Baur}}, \bibinfo {author}
  {\bibfnamefont {R.}~\bibnamefont {Bianchetti}}, \bibinfo {author}
  {\bibfnamefont {P.~J.}\ \bibnamefont {Leek}}, \bibinfo {author}
  {\bibfnamefont {A.}~\bibnamefont {Blais}}, \ and\ \bibinfo {author}
  {\bibfnamefont {A.}~\bibnamefont {Wallraff}},\ }\href@noop {} {\bibfield
  {journal} {\bibinfo  {journal} {Science}\ }\textbf {\bibinfo {volume}
  {322}},\ \bibinfo {pages} {1357} (\bibinfo {year} {2008})}\BibitemShut
  {NoStop}%
\bibitem [{\citenamefont {{Hanbury Brown}}\ and\ \citenamefont
  {Twiss}(1956)}]{Hanbury_Brown1956}%
  \BibitemOpen
  \bibfield  {author} {\bibinfo {author} {\bibfnamefont {R.}~\bibnamefont
  {{Hanbury Brown}}}\ and\ \bibinfo {author} {\bibfnamefont {R.~Q.}\
  \bibnamefont {Twiss}},\ }\href@noop {} {\bibfield  {journal} {\bibinfo
  {journal} {Nature (London)}\ }\textbf {\bibinfo {volume} {177}},\ \bibinfo
  {pages} {27} (\bibinfo {year} {1956})}\BibitemShut {NoStop}%
\bibitem [{\citenamefont {Kimble}\ \emph {et~al.}(1977)\citenamefont {Kimble},
  \citenamefont {Dagenais},\ and\ \citenamefont {Mandel}}]{Kimble1977}%
  \BibitemOpen
  \bibfield  {author} {\bibinfo {author} {\bibfnamefont {H.~J.}\ \bibnamefont
  {Kimble}}, \bibinfo {author} {\bibfnamefont {M.}~\bibnamefont {Dagenais}}, \
  and\ \bibinfo {author} {\bibfnamefont {L.}~\bibnamefont {Mandel}},\
  }\href@noop {} {\bibfield  {journal} {\bibinfo  {journal} {Phys. Rev. Lett.}\
  }\textbf {\bibinfo {volume} {39}},\ \bibinfo {pages} {691} (\bibinfo {year}
  {1977})}\BibitemShut {NoStop}%
\bibitem [{\citenamefont {Shapiro}\ \emph {et~al.}(1979)\citenamefont
  {Shapiro}, \citenamefont {Yuen},\ and\ \citenamefont {Mata}}]{Shapiro1979}%
  \BibitemOpen
  \bibfield  {author} {\bibinfo {author} {\bibfnamefont {J.}~\bibnamefont
  {Shapiro}}, \bibinfo {author} {\bibfnamefont {H.~P.}\ \bibnamefont {Yuen}}, \
  and\ \bibinfo {author} {\bibfnamefont {A.}~\bibnamefont {Mata}},\ }\href@noop
  {} {\bibfield  {journal} {\bibinfo  {journal} {IEEE Trans. Inf. Theory}\
  }\textbf {\bibinfo {volume} {25}},\ \bibinfo {pages} {179} (\bibinfo {year}
  {1979})}\BibitemShut {NoStop}%
\bibitem [{\citenamefont {Mandel}(1982)}]{Mandel1982}%
  \BibitemOpen
  \bibfield  {author} {\bibinfo {author} {\bibfnamefont {L.}~\bibnamefont
  {Mandel}},\ }\href@noop {} {\bibfield  {journal} {\bibinfo  {journal} {Phys.
  Rev. Lett.}\ }\textbf {\bibinfo {volume} {49}},\ \bibinfo {pages} {136}
  (\bibinfo {year} {1982})}\BibitemShut {NoStop}%
\bibitem [{\citenamefont {Slusher}\ \emph {et~al.}(1985)\citenamefont
  {Slusher}, \citenamefont {Hollberg}, \citenamefont {Yurke}, \citenamefont
  {Mertz},\ and\ \citenamefont {Valley}}]{Slusher1985}%
  \BibitemOpen
  \bibfield  {author} {\bibinfo {author} {\bibfnamefont {R.~E.}\ \bibnamefont
  {Slusher}}, \bibinfo {author} {\bibfnamefont {L.~W.}\ \bibnamefont
  {Hollberg}}, \bibinfo {author} {\bibfnamefont {B.}~\bibnamefont {Yurke}},
  \bibinfo {author} {\bibfnamefont {J.~C.}\ \bibnamefont {Mertz}}, \ and\
  \bibinfo {author} {\bibfnamefont {J.~F.}\ \bibnamefont {Valley}},\
  }\href@noop {} {\bibfield  {journal} {\bibinfo  {journal} {Phys. Rev. Lett.}\
  }\textbf {\bibinfo {volume} {55}},\ \bibinfo {pages} {2409} (\bibinfo {year}
  {1985})}\BibitemShut {NoStop}%
\bibitem [{\citenamefont {Smithey}\ \emph {et~al.}(1993)\citenamefont
  {Smithey}, \citenamefont {Beck}, \citenamefont {Raymer},\ and\ \citenamefont
  {Faridani}}]{Smithey1993}%
  \BibitemOpen
  \bibfield  {author} {\bibinfo {author} {\bibfnamefont {D.~T.}\ \bibnamefont
  {Smithey}}, \bibinfo {author} {\bibfnamefont {M.}~\bibnamefont {Beck}},
  \bibinfo {author} {\bibfnamefont {M.~G.}\ \bibnamefont {Raymer}}, \ and\
  \bibinfo {author} {\bibfnamefont {A.}~\bibnamefont {Faridani}},\ }\href@noop
  {} {\bibfield  {journal} {\bibinfo  {journal} {Phys. Rev. Lett.}\ }\textbf
  {\bibinfo {volume} {70}},\ \bibinfo {pages} {1244} (\bibinfo {year}
  {1993})}\BibitemShut {NoStop}%
\bibitem [{\citenamefont {Breitenbach}\ \emph {et~al.}(1997)\citenamefont
  {Breitenbach}, \citenamefont {Schiller},\ and\ \citenamefont
  {Mlynek}}]{Breitenbach1997}%
  \BibitemOpen
  \bibfield  {author} {\bibinfo {author} {\bibfnamefont {G.}~\bibnamefont
  {Breitenbach}}, \bibinfo {author} {\bibfnamefont {S.}~\bibnamefont
  {Schiller}}, \ and\ \bibinfo {author} {\bibfnamefont {J.}~\bibnamefont
  {Mlynek}},\ }\href@noop {} {\bibfield  {journal} {\bibinfo  {journal} {Nature
  (London)}\ }\textbf {\bibinfo {volume} {387}},\ \bibinfo {pages} {471}
  (\bibinfo {year} {1997})}\BibitemShut {NoStop}%
\bibitem [{\citenamefont {Silberhorn}(2007)}]{Silberhorn2007}%
  \BibitemOpen
  \bibfield  {author} {\bibinfo {author} {\bibfnamefont {C.}~\bibnamefont
  {Silberhorn}},\ }\href@noop {} {\bibfield  {journal} {\bibinfo  {journal}
  {Contemp. Phys.}\ }\textbf {\bibinfo {volume} {48}},\ \bibinfo {pages} {143}
  (\bibinfo {year} {2007})}\BibitemShut {NoStop}%
\bibitem [{\citenamefont {Auston}\ \emph {et~al.}(1984)\citenamefont {Auston},
  \citenamefont {Cheung},\ and\ \citenamefont {Smith}}]{Auston1984_2}%
  \BibitemOpen
  \bibfield  {author} {\bibinfo {author} {\bibfnamefont {D.~H.}\ \bibnamefont
  {Auston}}, \bibinfo {author} {\bibfnamefont {K.~P.}\ \bibnamefont {Cheung}},
  \ and\ \bibinfo {author} {\bibfnamefont {P.~R.}\ \bibnamefont {Smith}},\
  }\href@noop {} {\bibfield  {journal} {\bibinfo  {journal} {Appl. Phys.
  Lett.}\ }\textbf {\bibinfo {volume} {45}},\ \bibinfo {pages} {284} (\bibinfo
  {year} {1984})}\BibitemShut {NoStop}%
\bibitem [{\citenamefont {{Ch. Fattinger}}\ and\ \citenamefont
  {Grischkowsky}(1989)}]{Fattinger1989}%
  \BibitemOpen
  \bibfield  {author} {\bibinfo {author} {\bibnamefont {{Ch. Fattinger}}}\ and\
  \bibinfo {author} {\bibfnamefont {D.}~\bibnamefont {Grischkowsky}},\
  }\href@noop {} {\bibfield  {journal} {\bibinfo  {journal} {Appl. Phys.
  Lett.}\ }\textbf {\bibinfo {volume} {54}},\ \bibinfo {pages} {490} (\bibinfo
  {year} {1989})}\BibitemShut {NoStop}%
\bibitem [{\citenamefont {Auston}(1975)}]{Auston1975}%
  \BibitemOpen
  \bibfield  {author} {\bibinfo {author} {\bibfnamefont {D.~H.}\ \bibnamefont
  {Auston}},\ }\href@noop {} {\bibfield  {journal} {\bibinfo  {journal} {Appl.
  Phys. Lett.}\ }\textbf {\bibinfo {volume} {26}},\ \bibinfo {pages} {101}
  (\bibinfo {year} {1975})}\BibitemShut {NoStop}%
\bibitem [{\citenamefont {Wu}\ and\ \citenamefont {Zhang}(1995)}]{Wu1995}%
  \BibitemOpen
  \bibfield  {author} {\bibinfo {author} {\bibfnamefont {Q.}~\bibnamefont
  {Wu}}\ and\ \bibinfo {author} {\bibfnamefont {X.}~\bibnamefont {Zhang}},\
  }\href@noop {} {\bibfield  {journal} {\bibinfo  {journal} {Appl. Phys.
  Lett.}\ }\textbf {\bibinfo {volume} {67}},\ \bibinfo {pages} {3523} (\bibinfo
  {year} {1995})}\BibitemShut {NoStop}%
\bibitem [{\citenamefont {Nahata}\ \emph {et~al.}(1996)\citenamefont {Nahata},
  \citenamefont {Weling},\ and\ \citenamefont {Heinz}}]{Nahata1996}%
  \BibitemOpen
  \bibfield  {author} {\bibinfo {author} {\bibfnamefont {A.}~\bibnamefont
  {Nahata}}, \bibinfo {author} {\bibfnamefont {A.~S.}\ \bibnamefont {Weling}},
  \ and\ \bibinfo {author} {\bibfnamefont {T.~F.}\ \bibnamefont {Heinz}},\
  }\href@noop {} {\bibfield  {journal} {\bibinfo  {journal} {Appl. Phys.
  Lett.}\ }\textbf {\bibinfo {volume} {69}},\ \bibinfo {pages} {2321} (\bibinfo
  {year} {1996})}\BibitemShut {NoStop}%
\bibitem [{\citenamefont {Gallot}\ and\ \citenamefont
  {Grischkowsky}(1999)}]{Gallot1999}%
  \BibitemOpen
  \bibfield  {author} {\bibinfo {author} {\bibfnamefont {G.}~\bibnamefont
  {Gallot}}\ and\ \bibinfo {author} {\bibfnamefont {D.}~\bibnamefont
  {Grischkowsky}},\ }\href@noop {} {\bibfield  {journal} {\bibinfo  {journal}
  {J. Opt. Soc. Am. B}\ }\textbf {\bibinfo {volume} {16}},\ \bibinfo {pages}
  {1204} (\bibinfo {year} {1999})}\BibitemShut {NoStop}%
\bibitem [{\citenamefont {Liu}\ \emph {et~al.}(2004)\citenamefont {Liu},
  \citenamefont {Xu},\ and\ \citenamefont {Zhang}}]{Liu2004}%
  \BibitemOpen
  \bibfield  {author} {\bibinfo {author} {\bibfnamefont {K.}~\bibnamefont
  {Liu}}, \bibinfo {author} {\bibfnamefont {J.}~\bibnamefont {Xu}}, \ and\
  \bibinfo {author} {\bibfnamefont {X.-C.}\ \bibnamefont {Zhang}},\ }\href@noop
  {} {\bibfield  {journal} {\bibinfo  {journal} {Appl. Phys. Lett.}\ }\textbf
  {\bibinfo {volume} {85}},\ \bibinfo {pages} {863} (\bibinfo {year}
  {2004})}\BibitemShut {NoStop}%
\bibitem [{\citenamefont {K{\"{u}}bler}\ \emph {et~al.}(2004)\citenamefont
  {K{\"{u}}bler}, \citenamefont {Huber}, \citenamefont {T{\"{u}}bel},\ and\
  \citenamefont {Leitenstorfer}}]{Kuebler2004}%
  \BibitemOpen
  \bibfield  {author} {\bibinfo {author} {\bibfnamefont {C.}~\bibnamefont
  {K{\"{u}}bler}}, \bibinfo {author} {\bibfnamefont {R.}~\bibnamefont {Huber}},
  \bibinfo {author} {\bibfnamefont {S.}~\bibnamefont {T{\"{u}}bel}}, \ and\
  \bibinfo {author} {\bibfnamefont {A.}~\bibnamefont {Leitenstorfer}},\
  }\href@noop {} {\bibfield  {journal} {\bibinfo  {journal} {Appl. Phys.
  Lett.}\ }\textbf {\bibinfo {volume} {85}},\ \bibinfo {pages} {3360} (\bibinfo
  {year} {2004})}\BibitemShut {NoStop}%
\bibitem [{\citenamefont {Basov}\ \emph {et~al.}(2011)\citenamefont {Basov},
  \citenamefont {Averitt}, \citenamefont {van~der Marel}, \citenamefont
  {Dressel},\ and\ \citenamefont {Haule}}]{Basov2011}%
  \BibitemOpen
  \bibfield  {author} {\bibinfo {author} {\bibfnamefont {D.~N.}\ \bibnamefont
  {Basov}}, \bibinfo {author} {\bibfnamefont {R.~D.}\ \bibnamefont {Averitt}},
  \bibinfo {author} {\bibfnamefont {D.}~\bibnamefont {van~der Marel}}, \bibinfo
  {author} {\bibfnamefont {M.}~\bibnamefont {Dressel}}, \ and\ \bibinfo
  {author} {\bibfnamefont {K.}~\bibnamefont {Haule}},\ }\href@noop {}
  {\bibfield  {journal} {\bibinfo  {journal} {Rev. Mod. Phys.}\ }\textbf
  {\bibinfo {volume} {83}},\ \bibinfo {pages} {471} (\bibinfo {year}
  {2011})}\BibitemShut {NoStop}%
\bibitem [{\citenamefont {Ulbricht}\ \emph {et~al.}(2011)\citenamefont
  {Ulbricht}, \citenamefont {Hendry}, \citenamefont {Shan}, \citenamefont
  {Heinz},\ and\ \citenamefont {Bonn}}]{Ulbricht2011}%
  \BibitemOpen
  \bibfield  {author} {\bibinfo {author} {\bibfnamefont {R.}~\bibnamefont
  {Ulbricht}}, \bibinfo {author} {\bibfnamefont {E.}~\bibnamefont {Hendry}},
  \bibinfo {author} {\bibfnamefont {J.}~\bibnamefont {Shan}}, \bibinfo {author}
  {\bibfnamefont {T.~F.}\ \bibnamefont {Heinz}}, \ and\ \bibinfo {author}
  {\bibfnamefont {M.}~\bibnamefont {Bonn}},\ }\href@noop {} {\bibfield
  {journal} {\bibinfo  {journal} {Rev. Mod. Phys.}\ }\textbf {\bibinfo {volume}
  {83}},\ \bibinfo {pages} {543} (\bibinfo {year} {2011})}\BibitemShut
  {NoStop}%
\bibitem [{\citenamefont {Kienberger}\ \emph {et~al.}(2004)\citenamefont
  {Kienberger}, \citenamefont {Goulielmakis}, \citenamefont {Uiberacker},
  \citenamefont {Baltuska}, \citenamefont {Yakovlev}, \citenamefont {Bammer},
  \citenamefont {Scrinzi}, \citenamefont {Westerwalbesloh}, \citenamefont
  {Kleineberg}, \citenamefont {Heinzmann}, \citenamefont {Drescher},\ and\
  \citenamefont {Krausz}}]{Kienberger2004}%
  \BibitemOpen
  \bibfield  {author} {\bibinfo {author} {\bibfnamefont {R.}~\bibnamefont
  {Kienberger}}, \bibinfo {author} {\bibfnamefont {E.}~\bibnamefont
  {Goulielmakis}}, \bibinfo {author} {\bibfnamefont {M.}~\bibnamefont
  {Uiberacker}}, \bibinfo {author} {\bibfnamefont {A.}~\bibnamefont
  {Baltuska}}, \bibinfo {author} {\bibfnamefont {V.}~\bibnamefont {Yakovlev}},
  \bibinfo {author} {\bibfnamefont {F.}~\bibnamefont {Bammer}}, \bibinfo
  {author} {\bibfnamefont {A.}~\bibnamefont {Scrinzi}}, \bibinfo {author}
  {\bibfnamefont {T.}~\bibnamefont {Westerwalbesloh}}, \bibinfo {author}
  {\bibfnamefont {U.}~\bibnamefont {Kleineberg}}, \bibinfo {author}
  {\bibfnamefont {U.}~\bibnamefont {Heinzmann}}, \bibinfo {author}
  {\bibfnamefont {M.}~\bibnamefont {Drescher}}, \ and\ \bibinfo {author}
  {\bibfnamefont {F.}~\bibnamefont {Krausz}},\ }\href@noop {} {\bibfield
  {journal} {\bibinfo  {journal} {Nature (London)}\ }\textbf {\bibinfo {volume}
  {427}},\ \bibinfo {pages} {817} (\bibinfo {year} {2004})}\BibitemShut
  {NoStop}%
\bibitem [{\citenamefont {Planken}\ \emph {et~al.}(2001)\citenamefont
  {Planken}, \citenamefont {Nienhuys}, \citenamefont {Bakker},\ and\
  \citenamefont {Wenckebach}}]{Planken2001}%
  \BibitemOpen
  \bibfield  {author} {\bibinfo {author} {\bibfnamefont {P.~C.~M.}\
  \bibnamefont {Planken}}, \bibinfo {author} {\bibfnamefont {H.-K.}\
  \bibnamefont {Nienhuys}}, \bibinfo {author} {\bibfnamefont {H.~J.}\
  \bibnamefont {Bakker}}, \ and\ \bibinfo {author} {\bibfnamefont
  {T.}~\bibnamefont {Wenckebach}},\ }\href@noop {} {\bibfield  {journal}
  {\bibinfo  {journal} {J. Opt. Soc. Am. B}\ }\textbf {\bibinfo {volume}
  {18}},\ \bibinfo {pages} {313} (\bibinfo {year} {2001})}\BibitemShut
  {NoStop}%
\bibitem [{\citenamefont {Namba}(1961)}]{Namba1961}%
  \BibitemOpen
  \bibfield  {author} {\bibinfo {author} {\bibfnamefont {S.}~\bibnamefont
  {Namba}},\ }\href@noop {} {\bibfield  {journal} {\bibinfo  {journal} {J. Opt.
  Soc. Am.}\ }\textbf {\bibinfo {volume} {51}},\ \bibinfo {pages} {76}
  (\bibinfo {year} {1961})}\BibitemShut {NoStop}%
\bibitem [{Sup()}]{Suppl_Mat}%
  \BibitemOpen
  \href@noop {} {}\bibinfo {note} {See Supplemental Material at [URL will be
  inserted by publisher] for details.}\BibitemShut {Stop}%
\bibitem [{\citenamefont {New}(2011)}]{New_book}%
  \BibitemOpen
  \bibfield  {author} {\bibinfo {author} {\bibfnamefont {G.}~\bibnamefont
  {New}},\ }\href@noop {} {\emph {\bibinfo {title} {Introduction to Nonlinear
  Optics}}}\ (\bibinfo  {publisher} {Cambridge University Press},\ \bibinfo
  {address} {New York},\ \bibinfo {year} {2011})\BibitemShut {NoStop}%
\bibitem [{\citenamefont {Powers}(2011)}]{Powers_book}%
  \BibitemOpen
  \bibfield  {author} {\bibinfo {author} {\bibfnamefont {P.~E.}\ \bibnamefont
  {Powers}},\ }\href@noop {} {\emph {\bibinfo {title} {Fundamentals of
  Nonlinear Optics}}}\ (\bibinfo  {publisher} {Taylor {\&} Francis},\ \bibinfo
  {address} {Boca Raton},\ \bibinfo {year} {2011})\BibitemShut {NoStop}%
\bibitem [{\citenamefont {Yariv}(1989)}]{Yariv_book}%
  \BibitemOpen
  \bibfield  {author} {\bibinfo {author} {\bibfnamefont {A.}~\bibnamefont
  {Yariv}},\ }\href@noop {} {\emph {\bibinfo {title} {Quantum Electronics}}}\
  (\bibinfo  {publisher} {John Wiley \& Sons},\ \bibinfo {address} {New York},\
  \bibinfo {year} {1989})\BibitemShut {NoStop}%
\bibitem [{\citenamefont {Boyd}(2008)}]{Boyd_book}%
  \BibitemOpen
  \bibfield  {author} {\bibinfo {author} {\bibfnamefont {R.~W.}\ \bibnamefont
  {Boyd}},\ }\href@noop {} {\emph {\bibinfo {title} {Nonlinear Optics (Third
  Edition)}}}\ (\bibinfo  {publisher} {Academic Press},\ \bibinfo {address}
  {Burlington},\ \bibinfo {year} {2008})\BibitemShut {NoStop}%
\bibitem [{\citenamefont {Shen}(1984)}]{Shen_book}%
  \BibitemOpen
  \bibfield  {author} {\bibinfo {author} {\bibfnamefont {Y.~R.}\ \bibnamefont
  {Shen}},\ }\href@noop {} {\emph {\bibinfo {title} {Principles of nonlinear
  optics}}}\ (\bibinfo  {publisher} {Wiley-Interscience, New York},\ \bibinfo
  {year} {1984})\BibitemShut {NoStop}%
\bibitem [{\citenamefont {Allen}\ \emph {et~al.}(1992)\citenamefont {Allen},
  \citenamefont {Beijersbergen}, \citenamefont {Spreeuw},\ and\ \citenamefont
  {Woerdman}}]{Allen1992}%
  \BibitemOpen
  \bibfield  {author} {\bibinfo {author} {\bibfnamefont {L.}~\bibnamefont
  {Allen}}, \bibinfo {author} {\bibfnamefont {M.~W.}\ \bibnamefont
  {Beijersbergen}}, \bibinfo {author} {\bibfnamefont {R.~J.~C.}\ \bibnamefont
  {Spreeuw}}, \ and\ \bibinfo {author} {\bibfnamefont {J.~P.}\ \bibnamefont
  {Woerdman}},\ }\href@noop {} {\bibfield  {journal} {\bibinfo  {journal}
  {Phys. Rev. A}\ }\textbf {\bibinfo {volume} {45}},\ \bibinfo {pages} {8185}
  (\bibinfo {year} {1992})}\BibitemShut {NoStop}%
\bibitem [{\citenamefont {Calvo}\ \emph {et~al.}(2006)\citenamefont {Calvo},
  \citenamefont {Pic\'on},\ and\ \citenamefont {Bagan}}]{Calvo2006}%
  \BibitemOpen
  \bibfield  {author} {\bibinfo {author} {\bibfnamefont {G.~F.}\ \bibnamefont
  {Calvo}}, \bibinfo {author} {\bibfnamefont {A.}~\bibnamefont {Pic\'on}}, \
  and\ \bibinfo {author} {\bibfnamefont {E.}~\bibnamefont {Bagan}},\
  }\href@noop {} {\bibfield  {journal} {\bibinfo  {journal} {Phys. Rev. A}\
  }\textbf {\bibinfo {volume} {73}},\ \bibinfo {pages} {013805} (\bibinfo
  {year} {2006})}\BibitemShut {NoStop}%
\bibitem [{\citenamefont {Knight}\ and\ \citenamefont
  {Allen}(1983)}]{Knight_book}%
  \BibitemOpen
  \bibfield  {author} {\bibinfo {author} {\bibfnamefont {P.}~\bibnamefont
  {Knight}}\ and\ \bibinfo {author} {\bibfnamefont {L.}~\bibnamefont {Allen}},\
  }\href@noop {} {\emph {\bibinfo {title} {Concepts of Quantum Optics}}}\
  (\bibinfo  {publisher} {Pergamon Press, Oxford},\ \bibinfo {year}
  {1983})\BibitemShut {NoStop}%
\bibitem [{Note1()}]{Note1}%
  \BibitemOpen
  \bibinfo {note} {Confer Ref.~\cite {Raymer1995} for a case with a simpler
  transverse mode structure}\BibitemShut {NoStop}%
\bibitem [{Note2()}]{Note2}%
  \BibitemOpen
  \bibinfo {note} {The mixed terms were also neglected already in Eq.~\protect
  \textup {\hbox {\mathsurround \z@ \protect \normalfont (\ignorespaces \ref
  {Eq:Psz_t}\unskip \@@italiccorr )}}. The second order terms in $\delta
  \protect \mathaccentV {hat}05E{\protect \mathbf {E}}''$ ($\delta \protect
  \mathaccentV {hat}05E{E}''^\dagger _z\delta \protect \mathaccentV
  {hat}05E{E}''_z$ and $\delta \protect \mathaccentV {hat}05E{E}''^\dagger
  _s\delta \protect \mathaccentV {hat}05E{E}''_s$) do not contribute to the
  expectation value of the signal, neither to its variance or any higher
  moments.}\BibitemShut {Stop}%
\bibitem [{Note3()}]{Note3}%
  \BibitemOpen
  \bibinfo {note} {The phase shift is of no physical importance for the vacuum
  field contribution}\BibitemShut {NoStop}%
\bibitem [{\citenamefont {Brida}\ \emph {et~al.}(2014)\citenamefont {Brida},
  \citenamefont {Krauss}, \citenamefont {Sell},\ and\ \citenamefont
  {Leitenstorfer}}]{Brida2014}%
  \BibitemOpen
  \bibfield  {author} {\bibinfo {author} {\bibfnamefont {D.}~\bibnamefont
  {Brida}}, \bibinfo {author} {\bibfnamefont {G.}~\bibnamefont {Krauss}},
  \bibinfo {author} {\bibfnamefont {A.}~\bibnamefont {Sell}}, \ and\ \bibinfo
  {author} {\bibfnamefont {A.}~\bibnamefont {Leitenstorfer}},\ }\href@noop {}
  {\bibfield  {journal} {\bibinfo  {journal} {Laser Photon. Rev.}\ }\textbf
  {\bibinfo {volume} {8}},\ \bibinfo {pages} {409} (\bibinfo {year}
  {2014})}\BibitemShut {NoStop}%
\bibitem [{\citenamefont {Cingolani}\ \emph {et~al.}(1981)\citenamefont
  {Cingolani}, \citenamefont {Ferrara},\ and\ \citenamefont
  {Lugar{\`{a}}}}]{Cingolani1981}%
  \BibitemOpen
  \bibfield  {author} {\bibinfo {author} {\bibfnamefont {A.}~\bibnamefont
  {Cingolani}}, \bibinfo {author} {\bibfnamefont {M.}~\bibnamefont {Ferrara}},
  \ and\ \bibinfo {author} {\bibfnamefont {M.}~\bibnamefont {Lugar{\`{a}}}},\
  }\href@noop {} {\bibfield  {journal} {\bibinfo  {journal} {Solid State
  Commun.}\ }\textbf {\bibinfo {volume} {38}},\ \bibinfo {pages} {819 }
  (\bibinfo {year} {1981})}\BibitemShut {NoStop}%
\bibitem [{\citenamefont {Leitenstorfer}\ \emph {et~al.}(1999)\citenamefont
  {Leitenstorfer}, \citenamefont {Hunsche}, \citenamefont {Shah}, \citenamefont
  {Nuss},\ and\ \citenamefont {Knox}}]{Leitenstorfer1999}%
  \BibitemOpen
  \bibfield  {author} {\bibinfo {author} {\bibfnamefont {A.}~\bibnamefont
  {Leitenstorfer}}, \bibinfo {author} {\bibfnamefont {S.}~\bibnamefont
  {Hunsche}}, \bibinfo {author} {\bibfnamefont {J.}~\bibnamefont {Shah}},
  \bibinfo {author} {\bibfnamefont {M.~C.}\ \bibnamefont {Nuss}}, \ and\
  \bibinfo {author} {\bibfnamefont {W.~H.}\ \bibnamefont {Knox}},\ }\href@noop
  {} {\bibfield  {journal} {\bibinfo  {journal} {Appl. Phys. Lett.}\ }\textbf
  {\bibinfo {volume} {74}},\ \bibinfo {pages} {1516} (\bibinfo {year}
  {1999})}\BibitemShut {NoStop}%
\bibitem [{\citenamefont {Barnett}\ and\ \citenamefont
  {Radmore}(2002)}]{Barnett_book}%
  \BibitemOpen
  \bibfield  {author} {\bibinfo {author} {\bibfnamefont {S.}~\bibnamefont
  {Barnett}}\ and\ \bibinfo {author} {\bibfnamefont {P.}~\bibnamefont
  {Radmore}},\ }\href@noop {} {\emph {\bibinfo {title} {Methods in Theoretical
  Quantum Optics}}},\ Oxford Series in Optical and Imaging Sciences\ (\bibinfo
  {publisher} {Oxford University Press},\ \bibinfo {address} {New York},\
  \bibinfo {year} {2002})\BibitemShut {NoStop}%
\bibitem [{\citenamefont {Vogel}\ and\ \citenamefont
  {Welsch}(2006)}]{Vogel_book}%
  \BibitemOpen
  \bibfield  {author} {\bibinfo {author} {\bibfnamefont {W.}~\bibnamefont
  {Vogel}}\ and\ \bibinfo {author} {\bibfnamefont {D.}~\bibnamefont {Welsch}},\
  }\href@noop {} {\emph {\bibinfo {title} {Quantum Optics}}}\ (\bibinfo
  {publisher} {Wiley},\ \bibinfo {address} {Weinheim},\ \bibinfo {year}
  {2006})\BibitemShut {NoStop}%
\bibitem [{\citenamefont {Blow}\ \emph {et~al.}(1990)\citenamefont {Blow},
  \citenamefont {Loudon}, \citenamefont {Phoenix},\ and\ \citenamefont
  {Shepherd}}]{Blow1990}%
  \BibitemOpen
  \bibfield  {author} {\bibinfo {author} {\bibfnamefont {K.~J.}\ \bibnamefont
  {Blow}}, \bibinfo {author} {\bibfnamefont {R.}~\bibnamefont {Loudon}},
  \bibinfo {author} {\bibfnamefont {S.~J.~D.}\ \bibnamefont {Phoenix}}, \ and\
  \bibinfo {author} {\bibfnamefont {T.~J.}\ \bibnamefont {Shepherd}},\
  }\href@noop {} {\bibfield  {journal} {\bibinfo  {journal} {Phys. Rev. A}\
  }\textbf {\bibinfo {volume} {42}},\ \bibinfo {pages} {4102} (\bibinfo {year}
  {1990})}\BibitemShut {NoStop}%
\bibitem [{\citenamefont {Caves}(1981)}]{Caves1981}%
  \BibitemOpen
  \bibfield  {author} {\bibinfo {author} {\bibfnamefont {C.~M.}\ \bibnamefont
  {Caves}},\ }\href@noop {} {\bibfield  {journal} {\bibinfo  {journal} {Phys.
  Rev. D}\ }\textbf {\bibinfo {volume} {23}},\ \bibinfo {pages} {1693}
  (\bibinfo {year} {1981})}\BibitemShut {NoStop}%
\bibitem [{\citenamefont {Walls}\ and\ \citenamefont
  {Milburn}(2008)}]{Walls_book}%
  \BibitemOpen
  \bibfield  {author} {\bibinfo {author} {\bibfnamefont {D.}~\bibnamefont
  {Walls}}\ and\ \bibinfo {author} {\bibfnamefont {G.}~\bibnamefont
  {Milburn}},\ }\href@noop {} {\emph {\bibinfo {title} {Quantum Optics}}}\
  (\bibinfo  {publisher} {Springer},\ \bibinfo {address} {Berlin},\ \bibinfo
  {year} {2008})\BibitemShut {NoStop}%
\bibitem [{\citenamefont {Anderson}\ \emph {et~al.}(1997)\citenamefont
  {Anderson}, \citenamefont {McAlister}, \citenamefont {Raymer},\ and\
  \citenamefont {Gupta}}]{Anderson1997}%
  \BibitemOpen
  \bibfield  {author} {\bibinfo {author} {\bibfnamefont {M.~E.}\ \bibnamefont
  {Anderson}}, \bibinfo {author} {\bibfnamefont {D.~F.}\ \bibnamefont
  {McAlister}}, \bibinfo {author} {\bibfnamefont {M.~G.}\ \bibnamefont
  {Raymer}}, \ and\ \bibinfo {author} {\bibfnamefont {M.~C.}\ \bibnamefont
  {Gupta}},\ }\href@noop {} {\bibfield  {journal} {\bibinfo  {journal} {J. Opt.
  Soc. Am. B}\ }\textbf {\bibinfo {volume} {14}},\ \bibinfo {pages} {3180}
  (\bibinfo {year} {1997})}\BibitemShut {NoStop}%
\bibitem [{\citenamefont {Slusher}\ \emph {et~al.}(1987)\citenamefont
  {Slusher}, \citenamefont {Grangier}, \citenamefont {LaPorta}, \citenamefont
  {Yurke},\ and\ \citenamefont {Potasek}}]{Slusher1987}%
  \BibitemOpen
  \bibfield  {author} {\bibinfo {author} {\bibfnamefont {R.~E.}\ \bibnamefont
  {Slusher}}, \bibinfo {author} {\bibfnamefont {P.}~\bibnamefont {Grangier}},
  \bibinfo {author} {\bibfnamefont {A.}~\bibnamefont {LaPorta}}, \bibinfo
  {author} {\bibfnamefont {B.}~\bibnamefont {Yurke}}, \ and\ \bibinfo {author}
  {\bibfnamefont {M.~J.}\ \bibnamefont {Potasek}},\ }\href@noop {} {\bibfield
  {journal} {\bibinfo  {journal} {Phys. Rev. Lett.}\ }\textbf {\bibinfo
  {volume} {59}},\ \bibinfo {pages} {2566} (\bibinfo {year}
  {1987})}\BibitemShut {NoStop}%
\bibitem [{\citenamefont {Raymer}\ \emph {et~al.}(1995)\citenamefont {Raymer},
  \citenamefont {Cooper}, \citenamefont {Carmichael}, \citenamefont {Beck},\
  and\ \citenamefont {Smithey}}]{Raymer1995}%
  \BibitemOpen
  \bibfield  {author} {\bibinfo {author} {\bibfnamefont {M.~G.}\ \bibnamefont
  {Raymer}}, \bibinfo {author} {\bibfnamefont {J.}~\bibnamefont {Cooper}},
  \bibinfo {author} {\bibfnamefont {H.~J.}\ \bibnamefont {Carmichael}},
  \bibinfo {author} {\bibfnamefont {M.}~\bibnamefont {Beck}}, \ and\ \bibinfo
  {author} {\bibfnamefont {D.~T.}\ \bibnamefont {Smithey}},\ }\href@noop {}
  {\bibfield  {journal} {\bibinfo  {journal} {J. Opt. Soc. Am. B}\ }\textbf
  {\bibinfo {volume} {12}},\ \bibinfo {pages} {1801} (\bibinfo {year}
  {1995})}\BibitemShut {NoStop}%
\end{thebibliography}

\begin{thebibliography}{10}

\bibitem{Planken2001SSS}
P.~C.~M. Planken, H.-K. Nienhuys, H.~J. Bakker, and T. Wenckebach, J. Opt. Soc.
  Am. B {\bf 18},  313  (2001).

\bibitem{New_bookSSS}
G. New, {\em Introduction to Nonlinear Optics} (Cambridge University Press, New
  York, 2011).

\bibitem{Powers_bookSSS}
P.~E. Powers, {\em Fundamentals of Nonlinear Optics} (Taylor {\&} Francis, Boca
  Raton, 2011).

\bibitem{Yariv_bookSSS}
A. Yariv, {\em Quantum Electronics} (John Wiley \& Sons, New York, 1989).

\bibitem{Gallot1999SSS}
G. Gallot and D. Grischkowsky, J. Opt. Soc. Am. B {\bf 16},  1204  (1999).

\bibitem{Calvo2006SSS}
G.~F. Calvo, A. Pic\'on, and E. Bagan, Phys. Rev. A {\bf 73},  013805  (2006).

\bibitem{Allen1992SSS}
L. Allen, M.~W. Beijersbergen, R.~J.~C. Spreeuw, and J.~P. Woerdman, Phys. Rev.
  A {\bf 45},  8185  (1992).

\bibitem{Abramowitz_bookSSS}
M. Abramowitz and I. Stegun, {\em Handbook of Mathematical Functions: with
  Formulas, Graphs, and Mathematical Tables} (Dover, New York, 2012).

\bibitem{Loudon_bookSSS}
R. Loudon, {\em The Quantum Theory of Light} (Oxford University Press, New
  York, 2000).

\bibitem{Brida2014SSS}
D. Brida, G. Krauss, A. Sell, and A. Leitenstorfer, Laser Photon. Rev. {\bf 8},
   409  (2014).

\bibitem{Sellmeier1871SSS}
W. Sellmeier, Ann. Phys. (Berlin) {\bf 219},  272  (1871).

\bibitem{Marple1964SSS}
D.~T.~F. Marple, J. Appl. Phys. {\bf 35},  539  (1964).

\bibitem{Leitenstorfer1999SSS}
A. Leitenstorfer {\it et~al.}, Appl. Phys. Lett. {\bf 74},  1516  (1999).

\end{thebibliography}

%


\onecolumngrid
\clearpage

\setcounter{equation}{0}
\setcounter{figure}{0}
\setcounter{table}{0}
\setcounter{page}{1}

\makeatletter

\renewcommand{\theequation}{S\arabic{equation}}
\renewcommand{\thefigure}{S\arabic{figure}}
\renewcommand{\bibnumfmt}[1]{[S#1]}
\renewcommand{\citenumfont}[1]{S#1}

\setcounter{secnumdepth}{3}
\renewcommand\thesection{\arabic{section}}

\begin{center}
\textbf{\large \underline{Supplemental Material}}\\[0.5cm]
\textbf{\large Paraxial Theory of Direct
Electro-Optic Sampling of the Quantum Vacuum}
\end{center}

\begin{center} A.S. Moskalenko, C. Riek, D.V. Seletskiy, G. Burkard,
and A. Leitenstorfer
\end{center}

\section{Geometry of nonlinear mixing}
For the nonlinear mixing in the EOX, we select the
orientation of $\vec{E}_{\mathrm{p}}$ parallel to the $z$-axis.
This choice of the polarization direction of the probe field
ensures that the maximum signal is detected in the electro-optic
detection scheme for a copropagating classical THz electric field
polarized perpendicular to the probe electric field
\cite{Planken2001SSS}. In the experiment, adjustment is achieved by
rotation of the EOX around the $[110]$ axis for fixed, mutually
perpendicular polarization directions of the probe and detected
electric fields. Also, only one of
two possible polarization modes of the detected field [the one
perpendicular to the probe field, i.e. oriented parallel to the
unit vector $\vec{e}_s$ in Fig.~\ref{Fig:Geometry}(b)] contributes
to the signal in this geometry. There is no THz field generated by optical rectification of
the probe for this orientation of the EOX.

The second-order nonlinear mixing of the
probe field $\vec{E}_{\mathrm{p}}(t)$ with the detected THz field
$\vec{E}_{_\mathrm{THz}}(t)$ then induces nonlinear polarization in the
EOX with the following components
\cite{New_bookSSS,Powers_bookSSS,Yariv_bookSSS}:
\begin{align}
   P^{(2)}_x(t)&=4\epsilon_0d_{xyz}E_{_\mathrm{THz},y}(t)E_{\mathrm{p},z}(t)\;
   \label{Eq:Px_t}
\end{align}
and similarly for the $y$-component with the interchange of indices
$x\leftrightarrow y$ in  Eq.~\eqref{Eq:Px_t}. Here $\epsilon_0$ is
the vacuum permittivity. For the zincblende-type EOX we adopt as an example,
the tensor components $d_{xyz}\equiv \chi^{(2)}_{xyz}/2$ and
$d_{yxz}\equiv \chi^{(2)}_{yxz}/2$ are both equal to the same
constant denoted by $d_{_{36}}$
\cite{New_bookSSS,Powers_bookSSS,Yariv_bookSSS}. This coefficient is
related to the constant $r_{_{\!41}}$ used for the description of
the Pockels effect as $d_{_{36}}=-n^4 r_{_{\!41}}/4$. $n$
is the refractive index at the central frequency of the probe electric field.
For the following discussion, it is convenient to introduce
$d=4d_{_{36}}=-n^4 r_{_{\!41}}$. Writing Eq.~\eqref{Eq:Px_t} as an
instantaneous relation in the time domain we assume that the
frequencies $\Omega$ of the THz field are lower
than the frequencies $\omega$ of the probe field and that the
second-order nonlinear coefficient can be considered
constant in the frequency range determined
by the spectral width of the probe field. The frequency dependence
of the nonlinear coefficient can be easily included writing the
corresponding equations in the frequency domain, similar to
Ref.~\cite{Gallot1999SSS} but taking care of the particular
geometry. However, the discussion of the geometrical issues is
more concise in the time domain whereas the effect of the
frequency dependence of the nonlinear coefficient is finally not
significant in our case.

The nonlinear polarization induced in the $(110)$ plane is given
by
$\vec{P}^{(2)}=\frac{1}{\sqrt{2}}(P^{(2)}_y-P^{(2)}_x)\vec{e}_s$,
i.e. $P^{(2)}_z=0$ and
$P^{(2)}_s=\frac{1}{\sqrt{2}}(P^{(2)}_y-P^{(2)}_x)$, using the
unit vectors $\vec{e}_s$ and $\vec{e}_z$ as a basis in this plane.
Taking Eq.~\eqref{Eq:Px_t}, expressing also the components of the quantized THz
field in this basis and neglecting the effect of
quantum mechanical fluctuations of the probe beam on the induced
nonlinear polarization, i.e. assuming a sufficiently strong probe field,
we arrive at Eq.~\eqref{Eq:Psz_t}. Inclusion of the vacuum
contribution for the probe beam at this place would mean taking
into account mixed second-order corrections linearly dependent on
both the vacuum fluctuations of the probe field and on the probed
THz field. In our present consideration, we neglect such  terms since they
do not lead to significant effects.

\section{Laguerre-Gaussian modes and paraxial electromagnetic field
quantization} In electro-optic sampling, propagation of the
NIR probe beam through the EOX can be well described within the
paraxial approximation. The same approximation can be naturally
used to describe the sampled multi-THz quantum fields. The
corresponding expression for a quantized electric field within the
paraxial approximation was derived in Ref.~\cite{Calvo2006SSS}.
In free space, with the propagation axis selected as shown in
Fig.~\ref{Fig:Geometry}, it reads
\begin{equation}\label{EqSM:paraxial_quantization}
  \hat{\vec{E}}(\vec{r},t)=-i\sum_{\alpha,l,p}\int_0^\infty\!\!
  \mathrm{d}k \sqrt{\frac{\hslash\Omega}{4\pi\epsilon_0}}
\left[\vec{e}_\alpha\hat{a}_{\alpha,l,p}(k)e^{i(kr_{_{\|}}-\Omega
t)}\mathrm{LG}_{lp}(\vec{r}_{_\perp},r_{_{\|}};k)
-\mathrm{H.c.}\right]\;,
\end{equation}
where $\hat{a}_{\alpha,l,p}(k)$ denotes the annihilation operator
for a photon with absolute value of the wave vector $k$, frequency $\Omega=c_0k$,
orbital quantum numbers $l,p$, and polarization direction
$\vec{e}_\alpha$. The spatial mode functions are given by the
Laguerre-Gaussian (LG) modes
$\mathrm{LG}_{lp}(\vec{r}_{_\perp},r_{_{\|}};k)\equiv
\mathrm{LG}_{lp}(\rperp,\varphi,r_{_{\|}};k)$ which can be
written as \cite{Allen1992SSS,Calvo2006SSS}
\begin{equation}\label{EqSM:LG_modes}
  \mathrm{LG}_{lp}(\rperp,\varphi,r_{_{\|}};k)=\sqrt{\frac{2p!}{\pi(|l|+p)!}}\frac{1}{w(r_{_{\|}})}
  \left(\frac{\sqrt{2}r_{_\perp}}{w(r_{_{\|}})}\right)^{\!|l|}\!L_p^{|l|}\left(\frac{2 \rperp^2}{w^2(r_{_{\|}})}\right)
  \exp\left[-\frac{2\rperp^2}{w^2(r_{_{\|}})}+il\varphi+i\frac{k\rperp^2}{2\mathcal{R}(r_{_{\|}})}+i\Phi_\mathrm{G}(r_{_{\|}})\right].
\end{equation}
Here $w(r_{_{\|}})=w_0\sqrt{1+r_{_{\|}}^2/l_{\mathrm
R}^2(\Omega)}$ is the transverse mode radius at the longitudinal
position $r_{_{\|}}$ with $w_0$ being the waist size  of the
probe beam (mode radius at $r_{_{\|}}=0$) and $l_{\mathrm
R}(\Omega)=k w_0^2/2$ denoting the Rayleigh range of the beam at
given $k$. $\mathcal{R}(r_{_{\|}})=r_{_{\|}}\left[1+l_{\mathrm
R}^2(\Omega)/r_{_{\|}}^2\right]$ is the phase-front radius,
$\Phi_\mathrm{G}(r_{_{\|}})=-(2p+|l|+1)\arctan(r_{_{\|}}/w_0)$ is
the Gouy phase and $L_p^{|l|}(x)$ are the associated Laguerre
polynomials \cite{Abramowitz_bookSSS}. The LG modes are
normalized such that $\int_0^{2\pi}\!\mathrm{d}\phi
\int_0^{\infty}\!\mathrm{d}\rperp\rperp\;
\mathrm{LG}^*_{lp}(\rperp,\phi,r_{_{\|}};k)\mathrm{LG}_{l'p'}(\rperp,\phi,r_{_{\|}};k)=\delta_{ll'}\delta_{pp'}$
(for any $k$ and $r_{_{\|}}$), where $\delta_{ij}$ denotes the
Kronecker delta. The annihilation and creation operators satisfy
the continuum commutation relations
$[\hat{a}_{\alpha,l,p}(k),\hat{a}_{\alpha',l',p'}(k')]=[\hat{a}^\dagger_{\alpha,l,p}(k),\hat{a}^\dagger_{\alpha',l',p'}(k')]=0$
and
$[\hat{a}_{\alpha,l,p}(k),\hat{a}^\dagger_{\alpha',l',p'}(k')]=\delta_{\alpha\alpha'}\delta_{ll'}\delta_{pp'}\delta(k-k')$.
Expressing the creation and annihilation operators as functions of
frequency, whereby they satisfy
$[\hat{a}_{\alpha,l,p}(\Omega),\hat{a}_{\alpha',l',p'}(\Omega')]=[\hat{a}^\dagger_{\alpha,l,p}(\Omega),\hat{a}^\dagger_{\Omega',l',p'}(k')]=0$
and
$[\hat{a}_{\alpha,l,p}(\Omega),\hat{a}^\dagger_{\alpha',l',p'}(\Omega')]=\delta_{\alpha\alpha'}\delta_{ll'}\delta_{pp'}\delta(\Omega-\Omega')$,
Eq.~\eqref{EqSM:paraxial_quantization} transforms into
\begin{equation}\label{EqSM:paraxial_quantization_FD}
  \hat{\vec{E}}(\vec{r},t)=-i\sum_{\alpha,l,p}\int_0^\infty\!\!
  \mathrm{d}\Omega \sqrt{\frac{\hslash\Omega}{4\pi\epsilon_0c_0}}
\left[\vec{e}_\alpha\hat{a}_{\alpha,l,p}(\Omega)e^{i(k_\Omega
r_{_{\|}}-\Omega
t)}\mathrm{LG}_{lp}(\vec{r}_{_\perp},r_{_{\|}};k_{_\Omega})
-\mathrm{H.c.}\right]\;.
\end{equation}
By writing $k_{_\Omega}$ we have stressed that we consider
$k\equiv k_{_\Omega}=\Omega/c_0$ as a function of $\Omega$ in this
expression.
Note that the factor of $16\pi^3$ in the denominator under the square root in Eq.~(20) of  Ref.~\cite{Calvo2006SSS} needs to be replaced by $4\pi$.
%
%
This fact is
confirmed by deriving the expression for the total energy  operator
of the electro-magnetic field and has been considered in Eq.~\eqref{EqSM:paraxial_quantization}. From
Eq.~\eqref{EqSM:paraxial_quantization_FD}, the total energy
operator $\hat{\mathcal{E}}$ is obtained in its correct form as
$\hat{\mathcal{E}}=\int_0^\infty\!\mathrm{d}\Omega\;\hslash\Omega
\sum_{\alpha,l,p}
\hat{a}^\dagger_{\alpha,l,p}(\Omega)\hat{a}_{\alpha,l,p}(\Omega)$.

In media with refractive index $n_{_\Omega}$, the factor under the
square root in Eq.~\eqref{EqSM:paraxial_quantization_FD} should be
additionally divided by $n_{_\Omega}$
\cite[pp.~391-392]{Loudon_bookSSS}. This measure again ensures a correct
expression for the total energy operator of the field so that
Eq.~\eqref{EqSM:paraxial_quantization_FD} takes the form
\begin{equation}\label{EqSM:paraxial_quantization_FD_media}
  \hat{\vec{E}}(\vec{r},t)=-i\sum_{\alpha,l,p}\int_0^\infty\!\!
  \mathrm{d}\Omega \sqrt{\frac{\hslash\Omega}{4\pi\epsilon_0c_0n_{_\Omega}}}
\left[\vec{e}_\alpha\hat{a}_{\alpha,l,p}(\Omega)e^{i(k_{_\Omega}
r_{_{\|}}-\Omega
t)}\mathrm{LG}_{lp}(\vec{r}_{_\perp},r_{_{\|}};k_{_\Omega})
-\mathrm{H.c.}\right]\;.
\end{equation}

When we consider a thin EOX located at the beam waist
($r_{_{\|}}=0$), we use
$\mathrm{LG}_{lp}(\vec{r}_{_\perp},r_{_{\|}};k)\approx
\mathrm{LG}_{lp}(\vec{r}_{_\perp},r_{_{\|}}=0;k)\equiv
g_{lp}(\vec{r}_{_\perp})$, which are given by
\begin{equation}
  g_{lp}(\vec{r}_{_\perp})=\sqrt{\frac{2p!}{\pi(|l|+p)!}}\;
   \frac{1}{w_0}\left(\frac{\sqrt{2}\rperp}{w_0}\right)^{\!|l|}\!\!L^{|l|}_p\!\left(\frac{2\rperp^2}{w_0^2}\right)\exp\!\left(-\frac{\rperp^2}{w_0^2}+i
   l\varphi\right)\;.
\end{equation}
The transverse modes $g_{lp}(\vec{r}_{_\perp})\equiv
g_{lp}(\rperp,\phi)$ are independent of $k$. They are
normalized such that
\begin{equation*}
  \int_0^{2\pi}\!\mathrm{d}\phi
 \int_0^{\infty}\!\mathrm{d}\rperp\rperp\;
  g^*_{lp}(\rperp,\phi)g_{l'p'}(\rperp,\phi)=\delta_{l,l'}\delta_{p,p'}\;.
\end{equation*}
The fundamental (lowest-order) mode is Gaussian-shaped and
given by
\begin{equation}\label{EqSM:main_mode}
  g_{_{00}}(\vec{r}_{_\perp})=\sqrt{\frac{2}{\pi}}\frac{1}{w_0}\exp\!\left(-\frac{\rperp^2}{w_0^2}\right)\;.
\end{equation}
When the EOX is located at $r_{_{\|}}\neq 0$, the same
approximation applies and one can proceed similarly, just changing
$w_0\rightarrow w(r_{_{\|}})$ in the expressions for
$g_{lp}(\vec{r}_{_\perp})$.

\section{Refractive indices and response function}

\begin{figure}
  \includegraphics[width=0.5\linewidth]{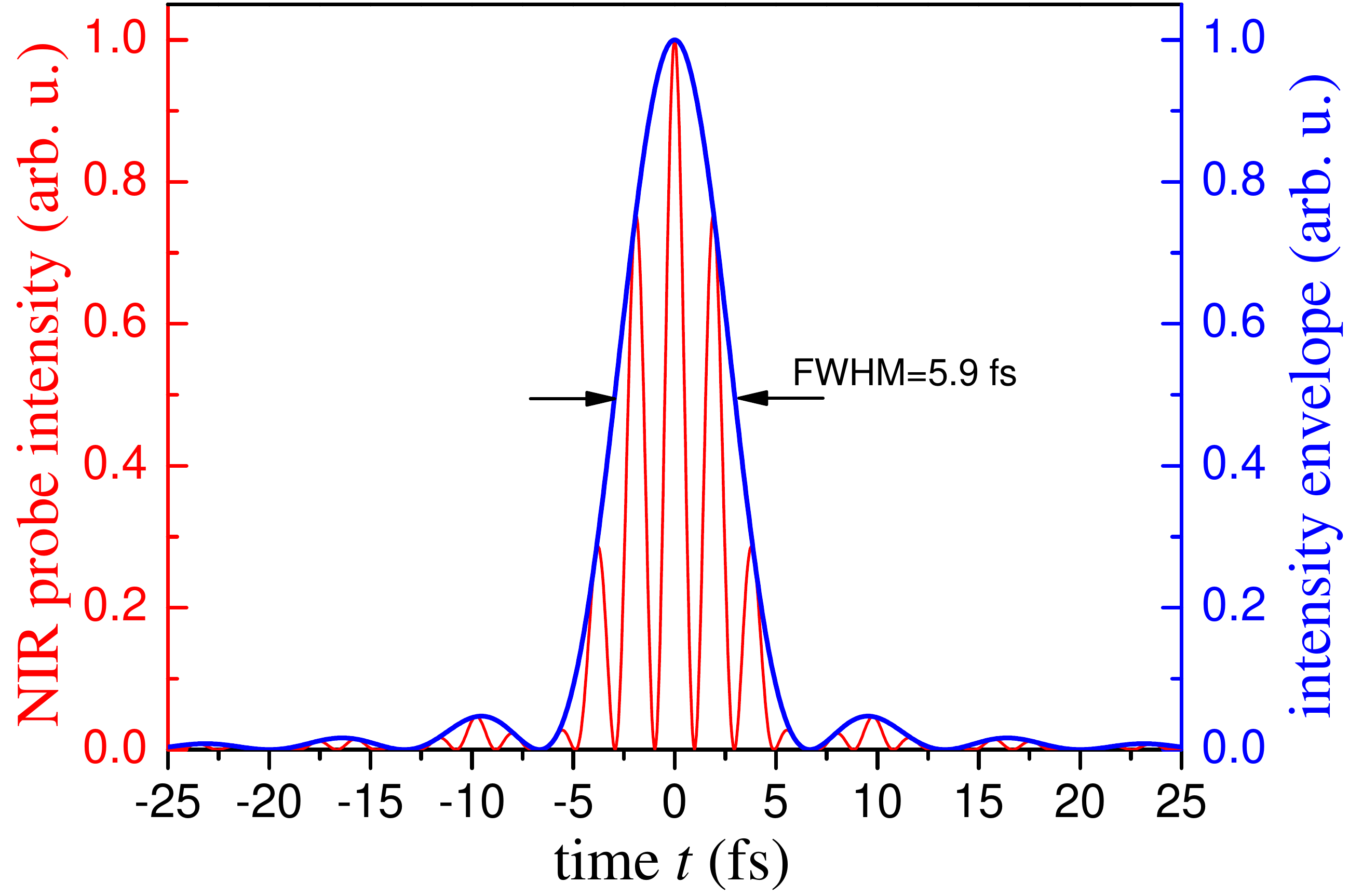}\\
  \caption{Temporal profile of the intensity (red line) and its envelope (blue line) of the NIR probe pulse used in our calculations.}\label{Fig:temp_profile}
\end{figure}

In the paper we used a sampling few-femtosecond NIR laser pulse of
the following specifications \cite{Brida2014SSS}: center frequency
$\omega_\mathrm{c}/(2\pi)=255~$THz, spectral bandwidth
$\Delta\omega/(2\pi)=150$ THz with rectangular spectral shape and
flat phase. The corresponding temporal profile of the NIR probe
intensity is shown in Fig.~\ref{Fig:temp_profile}. For such a
pulse we get $\omega_{\mathrm p}=247$~THz, where $\omega_{\mathrm
p}$ is defined in the text after Eq.~\eqref{Eq:S_EO2}. Notice that
$\omega_{\mathrm p}\approx \omega_\mathrm{c}$. However, there is a
small difference in these quantities due to different averaging
used in their definitions. This difference is of minor importance
for our consideration. The normalized Hermitian spectral
autocorrelation function $f(\Omega)$, as defined in the text after
Eq.~\eqref{Eq:S_EO2}, can be found as
\begin{equation}\label{EqSM:f_Omega}
    f(\Omega)=\left(1-\frac{|\Omega|}{\Delta\omega}\right)H(\Delta\omega-|\Omega|),
\end{equation}
where $H(x)$ denotes the Heaviside step function. For our example with a rectangular probe spectrum, $f(\Omega)$ takes
the shape of an isosceles triangle with the vertex at $\Omega=0$.
In order to determine the response function $R(\Omega)$, we have
to multiply $f(\Omega)$ by the phase-matching function
$\mathrm{sinc}\!\left[\frac{l\Omega}{2c_0}(n_{_\Omega}-n_\mathrm{g})\right]$.
The latter requires knowledge about the refractive index
$n_{_{_\Omega}}$ in the THz range and group refractive index
$n_\mathrm{g}$ at the (NIR) central frequency $\omega_\mathrm{c}$
of the probe pulse.

\begin{figure}
  \includegraphics[width=0.5\linewidth]{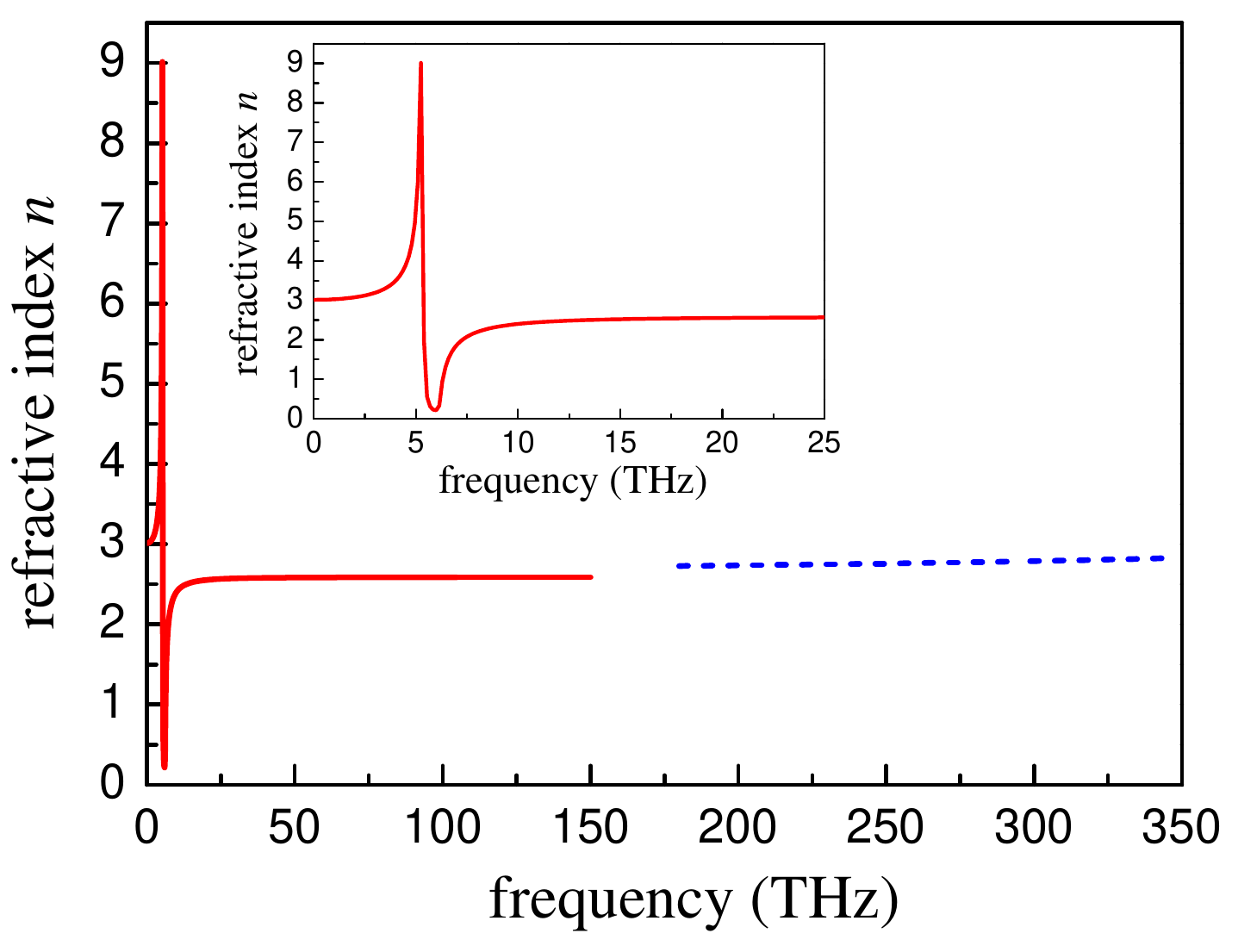}\\
  \caption{Dispersion of the refractive index in the THz (solid red line) and NIR (dashed blue line) range. Inset
  shows the dispersion with a higher frequency resolution in the range of small THz frequencies.}\label{Fig:n_omega}
\end{figure}

\begin{figure}
  \includegraphics[width=0.5\linewidth]{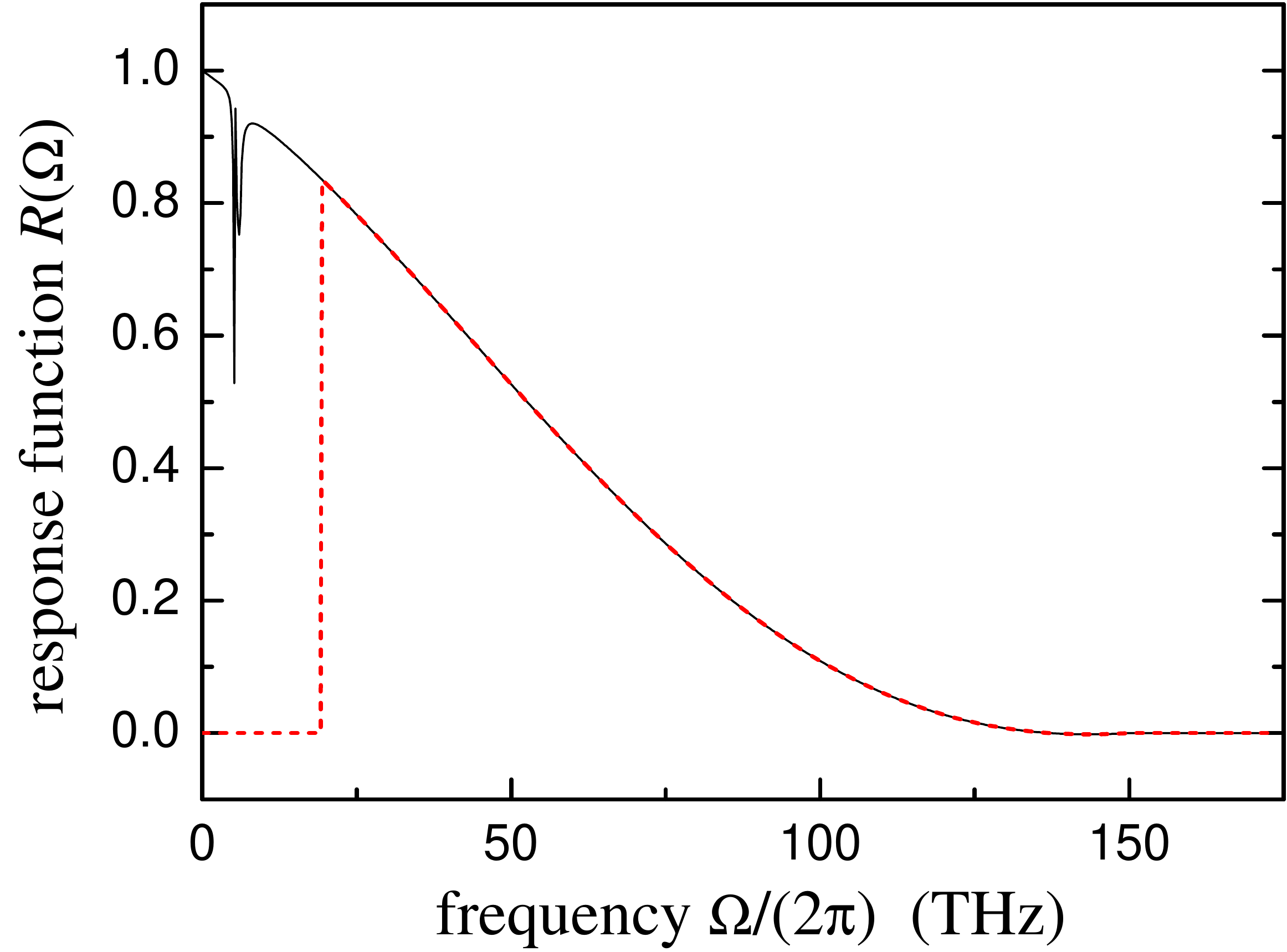}\\
  \caption{Calculated response function $R(\Omega)$
  without the low-frequency cutoff (solid black line) and with the
  low-frequency cutoff (dashed red line).}\label{Fig:R_function}
\end{figure}

Refractive index properties of a ZnTe crystal in the NIR frequency
range are modelled by a Sellmeier formula \cite{Sellmeier1871SSS}
\begin{equation}\label{EqSM:Sellmeir}
    n_\omega^2=A+[B\lambda^2/(\lambda^2-c^2)],
\end{equation}
with  $\lambda=2\pi c_0/\Omega$, $A=4.27$, $B=3.01$, and
$c^2=0.142$ \cite{Marple1964SSS}. The corresponding frequency
dependence is shown on the right side of Fig.~\ref{Fig:n_omega}.
For the refractive index $n_{_\Omega}$ in the THz frequency range,
we use the parametrization from Ref.~\cite{Leitenstorfer1999SSS}:
\begin{equation}\label{EqSM:n_Omega_THz}
    n_{_\Omega}=\mathrm{Re}\left(\sqrt{\left[1+\frac{(\hslash\omega_{_\mathrm{LO}})^2-(\hslash\omega_{_\mathrm{TO}})^2}
    {(\hslash\omega_{_\mathrm{TO}})^2-(\hslash\Omega)^2-i\hslash\gamma\Omega}\right]\epsilon_\infty}\right),
\end{equation}
with $\hslash\omega_{_\mathrm{TO}}=177$~cm$^{-1}$,
$\hslash\omega_{_\mathrm{LO}}=206$~cm$^{-1}$,
$\gamma=3.01$~cm$^{-1}$, and $\epsilon_\infty=6.7$. The
corresponding frequency dependence is shown on the left side of
Fig.~\ref{Fig:n_omega}. From these models we obtain the following
values of the refractive index and the group refractive index at
$\omega_\mathrm{c}=$255~THz: $n=2.76$ and $n_{\mathrm{g}}=2.9$. It
is important that these indices are almost constant in the
neighborhood of $\omega_\mathrm{c}$. Using the calculated
$n_{_\Omega}$ and $n_\mathrm{g}$ in the phase-matching function
with $l=7~\mu$m, the response function $R(\Omega)$ depicted
in Fig.~\ref{Fig:R_function} results. As discussed in the text of the
paper, we introduce a low-frequency cutoff excluding wavelengths
$\lambda$ with $\lambda/(2n_{_\Omega})>w_0$, in order to take into
account diffraction losses. The modified response
function is also shown in Fig.~\ref{Fig:R_function}. The resulting
integrand function entering the integral in
Eq.~\eqref{Eq:variance_THz_final} is found in
Fig.~\ref{Fig:signal_vacuum}(a) of the paper. Note that without
introducing the low-frequency cutoff we would get just a small
increase of approximately 21\% for the EOS variance calculated
from Eq.~\eqref{Eq:variance_THz_final}.

\section{EOS variance for the squeezed multi-THz vacuum}

For the EOS variance
$\langle\hat{\mathcal{S}}_\mathrm{eo}^2\rangle_\mathrm{sv}(\tau)$
normalized with respect to the EOS variance of the pure vacuum
$\langle\hat{\mathcal{S}}_\mathrm{eo}^2\rangle$, given by
Eq.~\eqref{Eq:variance_THz_final}, we obtain
\begin{equation}\label{EqSM:EOS_variance_squeezed}
    \langle\hat{\mathcal{S}}_\mathrm{eo}^2\rangle_\mathrm{sv}(\tau)/\langle\hat{\mathcal{S}}_\mathrm{eo}^2\rangle=
1\!+\!2aM\!+\!2b\sqrt{M(M+1)}\cos(\theta-2\Omega_\mathrm{c}\tau).
\end{equation}
Here  $M=\sinh r$, $a=I_a/I$, and $b=I_b/I$, where
\begin{eqnarray}
   I&=&\int_{0}^\infty\!\mathrm{d}\Omega\;\varrho^2(\Omega), \label{EqSM:I_coefficient}\\
   I_a&=&\int_{\Omega_1}^{\Omega_2}\!\mathrm{d}\Omega\;
\varrho^2(\Omega), \label{EqSM:Ia_coefficient}\\
   I_b&=&\int_{\Omega_1}^{\Omega_2}\!\mathrm{d}\Omega\;
\varrho(\Omega)\varrho(2\Omega_\mathrm{c}-\Omega),
\label{EqSM:Ib_coefficient}
\end{eqnarray}
with $\varrho(\Omega)=\sqrt{\Omega/n_{_\Omega}}R_0(\Omega)$. The
real coefficients $a$ and $b$ generally satisfy the
Cauchy-Bunyakovsky-Schwarz inequality $b\leq a$. Obviously, also
$a\leq 1$ is valid. The dependence of
$\langle\hat{\mathcal{S}}_\mathrm{eo}^2\rangle_\mathrm{sv}(\tau)/\langle\hat{\mathcal{S}}_\mathrm{eo}^2\rangle$
on the time delay $\tau$ is shown in Fig.~\ref{Fig:squeezing}(c)
for the same states as in Fig.~\ref{Fig:squeezing}(b) and sampling
parameters used for the pure vacuum case. We clearly see that for
certain delay times the EOS variance of the squeezed multi-THz
vacuum can beat the uncertainty limit set by the pure vacuum
state. Here, for $M\equiv\sinh r=2$, its minimum value constitutes
$\approx 36\%$ of the pure vacuum level. For the selected
parameters, a slightly stronger suppression of the quantum noise
down to $34\%$ can be achieved by increasing $M$ to $\approx5.8$,
whereas the maximal noise is more than doubled. These values
are generally determined by the coefficients $a$ and $b$,
introduced above, for which we have $b<a<1$ in the considered
case. Here, due to the limitation set by this inequality, a
complete noise suppression in the EOS variance is impossible for
any time delay.

\end{document}